\pdfoutput=1 
\documentclass{aa} 
\usepackage{longtable}
\usepackage{lscape}

\usepackage{graphicx}
\usepackage{txfonts}
\usepackage{natbib}
\usepackage{subfigure}
\usepackage{hyperref}
\usepackage{epstopdf}
\hypersetup{colorlinks=true,urlcolor=blue,citecolor=blue,pdfborder= 0 0 0}

\newcommand{\RNum}[1]{\uppercase\expandafter{\romannumeral #1\relax}}
\graphicspath{ {/home/jaanlaur/Documents/artiklid/2015 - OB} }
\bibpunct{(}{)}{;}{a}{}{,} 

\begin{document}

   \title{Variability survey of brightest stars in selected OB associations\thanks{Our variable star catalogue as well as light curves are only available at the CDS.}}
        \titlerunning{Variability survey of brightest stars}

   \author{Jaan Laur\inst{1,2}
          \and
          Indrek Kolka\inst{1}
                \and    
          T\~onis Eenm\"ae\inst{1}
                \and    
          Taavi Tuvikene\inst{1}
          \and    
          Laurits Leedjärv\inst{1}
          }

   \institute{
             Tartu Observatory, Observatooriumi~1, 61602 T\~oravere, Estonia\\
             \email{jaan.laur@to.ee}
          \and
             Institute of Physics, University of Tartu, W. Ostwaldi Str 1, 50411 Tartu, Estonia}
        
   \date{}

  \abstract
   {The stellar evolution theory of massive stars remains uncalibrated with high-precision photometric observational data mainly due to a small number of luminous stars that are monitored from space. Automated all-sky surveys have revealed numerous variable stars but most of the luminous stars are often overexposed. Targeted campaigns can improve the time base of photometric data for those objects.}
   {The aim of this investigation is to study the variability of luminous stars at different timescales in young open clusters and OB associations. }
   {We monitored 22 open clusters and associations from 2011 to 2013 using a 0.25-m telescope. Variable stars were detected by comparing the overall light-curve scatter with measurement uncertainties. Variability was analysed by the light curve feature extraction tool FATS. Periods of pulsating stars were determined using the discrete Fourier transform code SigSpec. We then classified the variable stars based on their pulsation periods and available spectral information.}
   {We obtained light curves for more than 20000 sources of which 354 were found to be variable. Amongst them we find 80 eclipsing binaries, 31 $\alpha$ Cyg, 13 $\beta$ Cep, 62 Be, 16 slowly pulsating B, 7 Cepheid, 1 $\gamma$ Doradus, 3 Wolf-Rayet and 63 late-type variable stars. Up to 55\% of these stars are potential new discoveries as they are not present in the Variable Star Index (VSX) database. We find the cluster membership fraction for variable stars to be 13\% with an upper limit of 35\%.}
   {}


   \maketitle

%

\section{Introduction}

The detailed stellar evolution theory of massive stars of over $8 M_{\odot}$ is currently an open question, as physical phenomena like magnetic fields, core convective overshooting, internal stellar rotation and angular momentum distribution inside stars are poorly understood.
There have been parametrised descriptions of those phenomena in corresponding models but they remain uncalibrated with high-precision observational data \citep{Aerts:2013}.
Concerning photometric monitoring, the space missions MOST, CoRoT and Kepler have observed hundreds of OB dwarfs during the last decade and the pulsational modelling of a few tens of objects has greatly contributed to the asteroseismic description of their internal structure.
But this is still an exploratory phase, and observations for ensembles of massive stars with known metallicity and age are needed \citep{Aerts:2015}.
The case of evolved OB stars is even more challenging, mainly due to the very small number of objects monitored from space up to now.
An additional obstacle is the relatively short campaign lengths ranging from weeks (MOST) to months (CoRoT; Kepler 2; BRITE) which is not sufficient to investigate longer oscillation periods reaching up to several months in the case of OB supergiants \citep[e.g.][]{Moravveji:2012}. 
The long-term high-duty-cycle monitoring is also useful for main-sequence variables ($\beta$~Cep and SPB stars) to check the possible long-term frequency or amplitude variability.

As a rule, ground-based photometric observations can not achieve the required precision of 0.1 to 0.5 mmag that is needed for proper asteroseismic analysis \citep{Aerts:2013}.
However, ground-based multicolour light curves (giving a precision of a few mmag) can be used to detect the pulsation modes with suitably high amplitudes \citep{Saesen:2013}.
For the less-studied supergiants, this approach would be especially profitable.

The spatial concentration of massive stars is highest in young open clusters and OB associations.
For example, $\sim$70\% of Galactic O-type stars are found in young clusters or loose OB associations \citep{Portegies:2010}, making these systems ideal targets for massive-star studies.
The scarcity of open clusters and associations older than 1 Gyr is attributed to encounters with giant molecular clouds, supported by the observed median cluster age in the solar neighbourhood of around 250 Myr \citep{Kharchenko:2005}.
This cluster age segregation means that different clusters contain massive stars in various evolutionary phases.
In this way, the targeted photometric monitoring of clusters and associations to get a high yield of massive stars will also provide an opportunity to study the possible relationship between the age and variability characteristics of selected objects.
In addition, the fraction of massive binary stars in bound systems can reach as high as $\sim~90~\pm~10\%$ when corrected for observational bias \citep{Kiminki:2012}.
Although photometric monitoring campaigns typically only detect binarity for system geometries that are close to edge on, each new eclipsing binary discovered among massive stars helps to constrain the binarity fraction.

It is not feasible to observe bright objects in large sky areas (e.g. clusters and associations) with high cadence using large-aperture telescopes because of their low availability and typically limited field of view.
But a number of small telescopes dedicated to astrophotographers exist that are suitable for bright-star observations with relatively large field of view and photometric filters.
Commercially available telescopes (e.g. iTelescope.net or slooh.com) give us access to astronomical sites with better astroclimates and minimal effort of maintaining the equipment.

The use of automated small telescopes has resulted in many variable-star all-sky surveys such as the All Sky Automated Survey \citep[ASAS,][]{Pojmanski:1997}, the Northern Sky Variability Survey \citep[NSVS,][]{Wozniak:2004}, and WASP \citep{Butters:2010} amongst others.
These surveys have revealed numerous new discoveries of variable stars \citep{Pojmanski:2005, Norton:2007, Nedoroscik:2015} from eclipsing binaries to stars with as yet undefined variability types.
But the need to cover nearly all of the observable sky takes its toll on the time resolution of the light curves for a given star, and therefore smaller-scope campaigns are needed to improve the observational time coverage for specific fields.
In addition, the brightest stars in the sky are often overexposed in those surveys and consequently understudied (bright limit for ASAS and NSVS is $V\approx8$ mag).

In order to study massive stars in different evolutionary phases, we photometrically observed a number of open clusters and OB associations in the northern sky over the 2011--2013 period with high cadence.
We used a small-aperture telescope with exposure times optimised for the brightest objects in those fields. 
Our basic goal is to investigate selected variable stars case-by-case and to incorporate additional spectroscopic data with appropriate spectral and temporal resolution \citep[e.g.][]{Laur:2015, Kolka:2013}. 
The goal of this paper is to expand the variable-star knowledge in our selected fields by extracting the variability types from the light curve information. 
Using a discrete Fourier transform method and time series feature extraction tool, we have analysed over 20 000 stars in our fields, and found 354 variable stars.
We describe the released catalogue and give an overview of the variable stars found in the data.

\section{Observations and data reduction}\label{sec:reduction}

\begin{table*}
\caption{Observed fields.}
\centering
\label{fields}
\begin{tabular}
{lccccrrrrr}
\hline \hline
\rule{0pt}{10pt}Field & RA & DEC & Nights & Cluster & \multicolumn{5}{c}{Cluster age} \\

 & Hms & Deg &    &  &\multicolumn{5}{c}{Myr}\\
 \hline 

\rule{0pt}{10pt}Berkeley 87 & 20:23:23  &+37:14:50& 363 & Berkeley 87 & $2.5^{(a)}$  &$14^{(b)}$& $13^{(c)}$ &$14^{(d)} $ &\\

\rule{0pt}{10pt}& && & MWSC 3336 & $30^{(c)}$&&&& \\

\rule{0pt}{10pt}NGC 6913 & 20:24:00& +38:15:43& 333 &  NGC 6913   &$5^{(b)}$& $32^{(c)}$&$13^{(d)}$ & $5^{(e)}$& $1.75^{(f)}$\\

\rule{0pt}{10pt}Cyg OB2 &20:32:41 &+41:22:00& 359& Cygnus OB2 & $5^{(c)}$& $1^{(g)}$& $1-7^{(h)}$&&\\

\rule{0pt}{10pt}& && & FSR 0236 & $158^{(b)}$& $158^{(c)}$&&&\\

\rule{0pt}{10pt}Berkeley 86 & 20:18:57 &+38:44:43& 231& Berkeley 86 & $3^{(a)}$& $13^{(b)}$& $6^{(c)}$&&\\

\rule{0pt}{10pt}P Cygni  & 20:18:56 &+38:02:09& 218 & IC 4996   & $7^{(b)}$ & $14^{(c)}$ & $9^{(d)}$& $10^{(i)}$&\\

\rule{0pt}{10pt}& && & Dolidze 42 & $35^{(b)}$&$33^{(c)}$ & $35^{(d)}$&&\\

\rule{0pt}{10pt}& && & Berkeley 85 & $1000^{(b)}$  & $1160^{(c)}$  & $1000^{(i)}$&&\\

\rule{0pt}{10pt}NGC 7510 & 23:14:30 &+60:30:30&240 & NGC 7510 &$22^{(b)}$&$50^{(c)}$&$38^{(d)}$ &$10^{(j)}$ &$6^{(k)}$\\

\rule{0pt}{10pt}& && & Markarian 50   &$12^{(b)}$  &$13^{(c)}$   &$12^{(d)}$   &$7.5^{(l)}$&\\

\rule{0pt}{10pt}& && & FSR 0422 & $461^{(b)}$& $461^{(c)}$&&&\\

\rule{0pt}{10pt}NGC 7654 & 23:25:24 &+61:23:40& 244&   NGC 7654  &$158^{(b)}$ &$79^{(c)}$ &$58^{(d)}$&$10^{(i)}$ &$100^{(m)}$\\

\rule{0pt}{10pt}& && & Czernik 43  & $40^{(b)}$ & $58^{(c)}$ & $50^{(d)}$ &&\\

\rule{0pt}{10pt}PZ Cas   & 23:45:52 &+61:56:00& 222& Stock 17 & $6^{(b)}$ & $10^{(c)}$ & $4-25^{(n)}$&&\\

\rule{0pt}{10pt}NGC 581 & 01:35:18 &+60:49:30& 160& NGC 581  & $22^{(b)}$ & $28^{(c)}$ & $16^{(o)}$&&\\

\rule{0pt}{10pt}NGC 663 & 01:45:37 &+61:12:00& 153& NGC 663  &$25^{(b)}$ &$32^{(c)}$ &$20^{(p)}$ &$13^{(q)}$ &$20-25^{(r)}$\\

\rule{0pt}{10pt}NGC 869/884   & 02:20:07 &+57:07:55& 158& NGC 869  &$12^{(b)}$ &$19^{(c)}$ &$11^{(s)}$ &$13.5^{(t)}$& \\

\rule{0pt}{10pt}&  && & NGC 884  &$13^{(b)}$ &$16^{(c)}$ &$11^{(s)}$ &$14^{(t)}$ &\\

\rule{0pt}{10pt}NGC 957 &02:30:17 &+57:39:30 & 131& NGC 957 &$10^{(b)}$ &$18^{(c)}$ &$11^{(d)}$ &$10^{(u)}$&\\

\rule{0pt}{10pt}IC 1805 & 02:33:00 &+61:33:00& 127& IC 1805  &$3^{(b)}$ &$18^{(c)}$ &$7^{(d)}$ &$1-7^{(s)}$ &$1-3^{(v)}$\\

\rule{0pt}{10pt}EO Per & 02:53:45&+57:36:00& 123&  --  & \multicolumn{5}{c}{}\\

\rule{0pt}{10pt}Gem OB1 &06:11:00 &+23:02:00& 159& Gemini OB1 & $9^{(r)}$&&&&\\

\hline
\end{tabular}
\tablebib{
(a) \citet{Massey:2001};
(b) \citet{Dias:2002} \footnote{\url{http://www.wilton.unifei.edu.br/ocdb/}};
(c) MWSC \citet{Kharchenko:2013} \footnote{\url{https://heasarc.gsfc.nasa.gov/W3Browse/all/mwsc.html}};
(d) WEBDA \footnote{\url{http://www.univie.ac.at/webda/navigation.html}};
(e) \citet{straizys:2014};
(f) \citet{Joshi:1983};
(g) \citet{Massey:1995};
(h) \citet{Wright:2014};
(i) \citet{Maciejewski:2007};
(j) \citet{Barbon:1996};
(k) \citet{Piskunov:2004};
(l) \citet{Baume:2004};
(m) \citet{Choi:1999};
(n) \citet{Pandey:1986};
(o) \citet{Sanner:1999};
(p) \citet{Georgy:2014}; 
(q) \citet{Pandey:2005};
(r) \citet{Pigulski:2001};
(s) \citet{Tetzlaff:2010} \footnote{\url{http://mnras.oxfordjournals.org/content/402/4/2369/T14.expansion.html}};
(t) \citet{Currie:2010};
(u) \citet{Yadav:2008};
(v) \citet{Wolff:2011}.
}
\end{table*}

We observed 22 northern open clusters and associations in four different constellations across 15 fields. 
The fields were chosen for their number of bright OB stars and the age difference of the clusters.
The observations were not centred on the clusters but instead the field of view was set to maximise the number of massive stars in each field based on the catalogue of supergiants and O stars in associations and clusters \citep{Humphreys:1978}.

The data set was obtained between 2011 and 2013 with the commercial iTelescope.net 0.25-m Takahashi Epsilon telescope (T4), located in Mayhill, New Mexico, USA.
An SBIG ST-10 CCD camera with Johnson-Cousins $BV\!I$ filters from Custom Scientific was used, providing a field of view of $60.5~\times~40.8$ arcmin with pixel scale of 1.64 arcsec. 

The time resolution was one of the priorities of the campaign and therefore observations were carried out as frequently as possible.
During the campaign, the telescope was used on 488 nights out of 996, having an uptime of 49\%.
We observed every available field once per night with occasional nights dedicated to specific fields with multiple observations.

A single observation consisted of three consecutive exposures in a set of $V\!I$ or $BV\!I$ passbands with exposure times optimised for the brightest sources in each field.
For some fields, we opted for two different exposure times in order not to saturate the brightest stars and, at the same time, to monitor some of the fainter massive stars in that field.
Analysis of our data set was mostly done using the $V$ passband.
$B$ passband exposures were only obtained for two of the fields, so we can not use it for consistency reasons, whereas $I$ passband exposures were optimised for the red supergiants, leaving most of the hotter sources underexposed and thus having larger scatter than in $V$ passband.

Table~\ref{fields} lists all of the observed fields, the number of nights they were monitored, and open clusters within those fields.
The ages of the clusters have been gathered from literature with the references given in parentheses.
In addition to smaller works, two larger catalogues were used for the cluster age determination.
The first is by \citet{Kharchenko:2013} who compiled a catalogue of over 470 million stars and use a homogeneous method for determining cluster parameters for 3 006 open clusters which are published as the Milky Way Star Clusters Catalogue (MWSC).
The other compilation of cluster data is the DAML02 \citep{Dias:2002} database, containing cluster parameters for more than 2 167 open clusters.
On a side note, the age discrepancy of NGC 7654 is thought to be caused by multiple and independent star-formation periods as there seems to be an older cluster behind a group of younger stars \citep{Pandey:2001}.

Data reduction was carried out using the PHOTWORK software \citep{Tuvikene:2012} based on the IDL Astronomy User's Library\footnote{\url{http://idlastro.gsfc.nasa.gov/}} and the magnitudes were extracted by means of aperture photometry.
We used an aperture size that was scaled by the full width at half maximum (FWHM) of stars in any given frame with the typical aperture radius of $1.8 \times$ FWHM.
For further analysis, we only used instrumental magnitudes with uncertainties less than 0.1 mag.

Additionally, SExtractor source extraction software \citep{Bertin:1996} was used for PSF photometry.
This was necessary to disentangle the light curves of stars where the signal of a source in a given aperture was contaminated by neighbouring stars.
We determined this contamination from FWHM-magnitude plots by a detection of a strong correlation between the FWHM and magnitude values.
This correlation indicates a nearby star that affects the total extracted magnitude when the aperture size changes with the FWHM value.
To disentangle these contaminated sources, PSF photometry was used.

We used Bouguer's extinction plots to flag data points affected by bad weather \citep{Sterken:1992}.
We flagged data points for which the brightness drop due to extra atmospheric extinction exceeded 0.8 magnitudes from the fitted magnitude to airmass slope.
This indicates that the bigger than average extinction in the atmosphere is due to clouds.

Flat-field frames were supplied by the service provider for between one and ten nights per month, with three to six flat frames per night.
Master flats were constructed individually for each observing night using raw flats from the ten closest nights.
We excluded frames that exhibited systematic differences over 5\% from the median flat of neighbouring nights. 
Typically between six and twelve frames from between two and four nights were combined to an individual master flat.
During periods with very few provided flats, we constructed master flats from frames of a single night with minimal time separation from the observations.

Scattered light reaching the focal plane of the camera causes illumination gradients in flat-field frames and using these flat fields in the data reduction may yield position-dependent systematic uncertainties in the resulting photometry \citep{Manfroid:1995}.
The T4 telescope had a German equatorial mount and due to the aforementioned scattered light problem, the data show systematic effects depending on the telescope being on the east or west side of the pier.
To correct for these effects we constructed correction surfaces for the full frame.
This was done by dividing the data of constant stars into distinct time intervals based on the systematics in the light curves and then measuring the offset of mean magnitudes relative to the first time interval.
In this way we found a global difference surface over the whole frame for each of these time intervals in every filter.
In total, we constructed correction surfaces for 14 time intervals that were separated based on the combination from the pierside position and other systematics in the data.
We generated two different types of correction surfaces using either a median smoothing or a polynomial fitting.
The analysis of light curves were done using a combination of both corrections.

The amount of flat fielding error increases with the distance from the centre of the frame.
So in order to minimise correction uncertainties from comparison stars, they were selected as close to the frame centre as possible.
The correction was applied to both the programme star and the comparison stars; the mean correction value is 0.017 magnitudes.
The nightly mean precision of brightest stars is better than the mean correction value (cf. Fig.~\ref{scatter_mag}) and as a result, the total signal-to-noise ratio of the data can be improved by implementing this correction.

A separate correction was carried out between Julian dates (JD) 2456189 and 2456243 as the whole field was rotated about 30 degrees.
Also, the period between JD 2456350 and 2456413 needed to be corrected as there were erroneous flat-field frames supplied.

\begin{figure}[t]
        \includegraphics{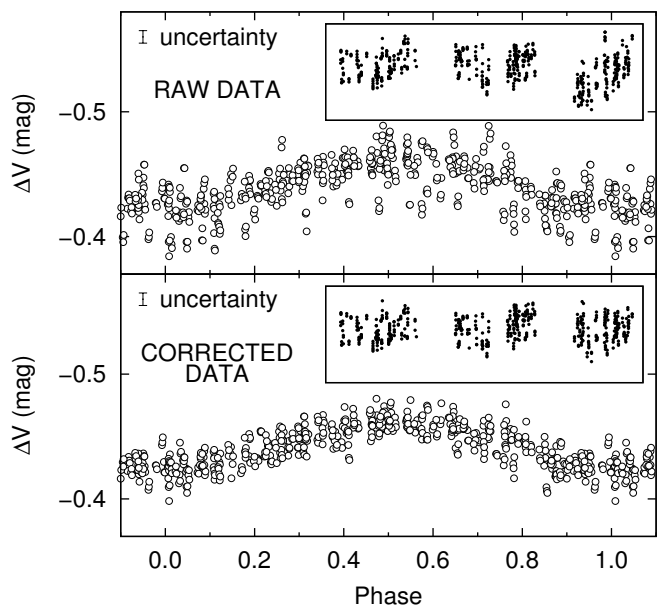}
        \caption{Example light curve correction for SPB star BD+59 2692. The light curve is folded with a period of 1.3032 days. The upper panel is the raw data and the lower panel is the final data that are obtained after using the correction surfaces. The main panels show a folded phase-curve and the small panels show light curves in time domain before and after correction.}
        \label{correction_example}
\end{figure}

An example correction of a light curve is given in Fig.~\ref{correction_example}.
The raw data (upper panel) show offsets in the light curve (inset panels) that are corrected (lower panel) based on the surfaces combined from constant stars.
The need for the correction is more evident in the phased light curves (big panels) where the scatter of a phase curve can be seen reduced after applying the correction.

Due to telescope control software issues, telescope centring on the target was also not consistent.
These issues caused the frames to be at offset with respect to the targeted field.
In the worst case, that offset was up to 1/4 of frame size, excluding about one third of the target stars.
However, this problem affected only a limited number of exposures.

To analyse the data, we used multiple comparison stars that were chosen to be, preferably, close to the centre of the frame to minimise systematic errors across the frame.
They were also chosen to be usable in both the$ V$ and $I$ passbands. 
The final comparison stars were selected by studying star pairs that matched the aforementioned criteria and provided minimal scatter in the combined light curve.
We used between two and five comparison stars per field. 
This combination of comparison stars gives us a relative photometric precision of  better than 0.01 mag.

In addition to the photometric observations, the classification of selected brighter variable stars without known spectral types was carried out using the 1.5 meter AZT-12 telescope at Tartu Observatory, Estonia.
Spectral data were collected using a long slit spectrograph ASP-32 with 600 lines/mm grating, yielding $\textnormal{R}\approx1550$ and covering wavelength range $\lambda \approx3680$ to $5830~\AA$.

All of the collected spectra were reduced using IRAF\footnote{\url{http://iraf.noao.edu/}} tasks.
First, a zero-level correction was applied along with the pixel-to-pixel non-uniformity corrections using incandescent lamp flat-field data. 
Then, for the wavelength calibration, a dispersion relation was established using a ThAr spectral lamp. 
Finally, to correct for the instrumental sensitivity effects, target spectra were corrected using one or more spectrophotometric standard stars from CALSPEC\footnote{\url{http://www.stsci.edu/hst/observatory/crds/calspec.html}} database \citep{Bohlin:2014} that were observed on the same nights as the targets.
For classification purposes, a digital spectral classification atlas\footnote{\url{http://ned.ipac.caltech.edu/level5/Gray/Gray_contents.html}} by R. O. Gray for late-type stars and synthetic spectra for B-type stars\footnote{\url{https://www.lsw.uni-heidelberg.de/projects/hot-stars/websynspec.php}} \citep{Gummersbach:1996} were used.

\section{Calibrated data}\label{sec:database}

\begin{figure}[t]
        \includegraphics[width=0.49\textwidth, trim=0 0 0 0,clip]{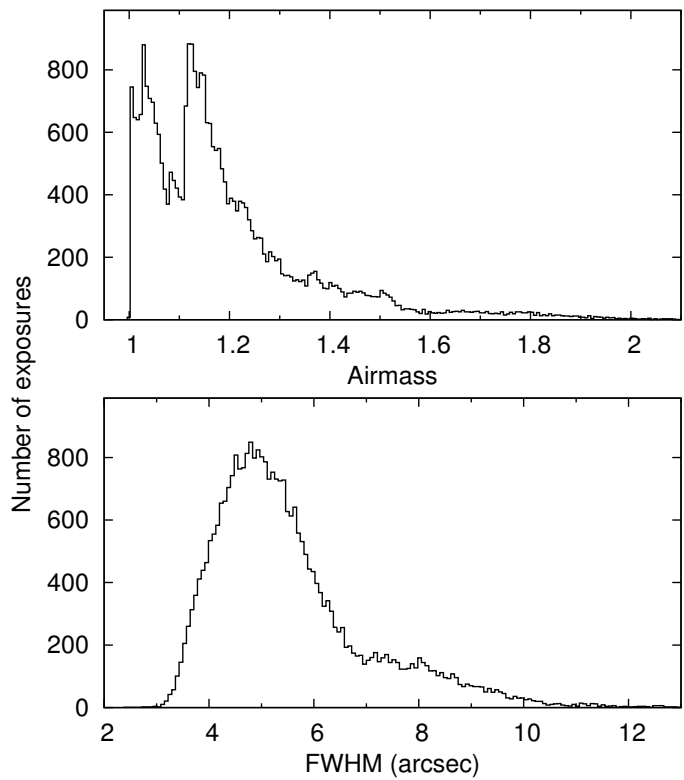}
        \caption{Histogram of the airmass values (upper panel) and FWHM values (lower panel) over the whole campaign.}
        \label{seeing_airmass}
\end{figure}

In total, 28 082 CCD frames were obtained and calibrated, of which 1 942 were obtained in $B$, 13 428 in the $V$ passband and 12 712 in the $I$ passband.

We give the airmass histogram in the upper panel of Fig.~\ref{seeing_airmass}.
As the observation plans for the campaign were sequenced to observe every field preferably at lower airmasses, the airmass values are mainly in the range of 1 to 1.5.
The two peaks of airmass values are due to two distinct maximum altitude ranges for our selected fields as seen from the New Mexico site.

The FWHM values on the lower panel in Fig.~\ref{seeing_airmass} were obtained from the CCD frames by analysing the brightest stars.
These values varied from 3 to 14 arcsec with the median value of 5.2 arcsec, affected by telescope optics.
Our obtained FWHM was good for observing bright stars, as we could use longer exposure times without overexposing our target stars.
Although larger FWHM lowers the signal-to-noise ratio of faint stars and affects our limiting magnitude, the focus was on the brightest cluster members and the trade-off was deemed acceptable.
We limited our analysis to using FWHM values of less than 10 arcsec, as the data become much worse at high values.

\begin{figure}[t]
        \includegraphics{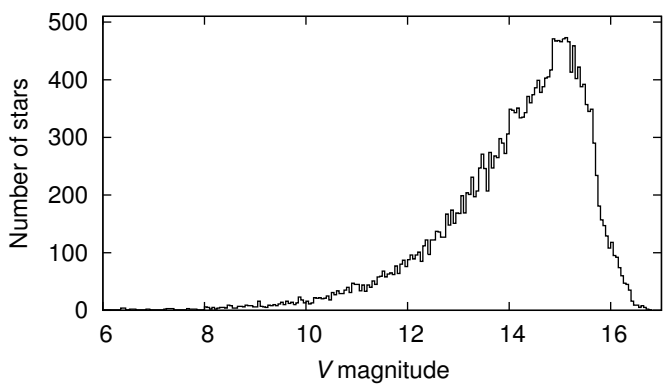}
        \caption{$V$ magnitude distribution of all the detected sources in our data set.}
        \label{Vmag}
\end{figure}

We obtained light curves for 25 395 sources.
Fig.~\ref{Vmag} shows the $V$ magnitude distribution of our observed stars.
The measured stellar brightnesses were not transformed into standard photometric system.
Instead we used an instrumental system that was shifted close to Johnson-Cousins system by using known standard magnitudes from the APASS\footnote{\url{https://www.aavso.org/apass}} DR9 catalogue.
The fall-off in the number of stars with the apparent magnitudes fainter than 15 mag is due to the short exposure times.

\addtolength{\tabcolsep}{-1pt} 

\begin{table}
\caption{Photometric quality of the $V$ passband exposures.}
\centering
\label{field_stars}
\begin{tabular}{lrrcrrr}
\hline \hline

\rule{0pt}{10pt}Field & Exp. & N\,\,\,       & $V$ range & Lim.\,      & $\sigma$\,\,\,\,\,         & N$_{90}$ \\
 & (s)\,\, &        & (mag)  &  (mag)     & (mag)         & \\
  \hline 

\rule{0pt}{10pt}Berkeley 87 & 60 &1509 & 6--16.5 & 14.0& 0.046& 254\\
NGC 6913 & 60 &1650 & 8--16.5 & 14.0 & 0.047& 265\\
Cyg OB2 & 60 & 1392 & 6--16.5 & 14.0& 0.045& 284\\
Berkeley 86 & 5/20 & 2033 & 8--15.5 & 12.0& 0.044& 102\\
P Cygni & 10 & 1311 & 7--15 & 12.5& 0.047& 98\\

NGC 7510 & 90 & 2146 & 8--17 & 14.0& 0.044& 456\\
NGC 7654 & 10/40 & 2777 & 6--16 & 12.5& 0.046 & 143\\
PZ Cas & 30 & 1946 & 6--16 & 13.5& 0.047& 225\\

NGC 581 & 8 & 850 & 7--14.5 & 12.0& 0.041& 95\\
NGC 663 & 40 & 1841 & 6--16 & 13.5& 0.043& 288\\
NGC 869/884 & 8/40 & 3321 & 6--16 & 12.5& 0.049& 269\\
NGC 957 & 20 & 1103 & 7--15 & 13.0& 0.045& 102\\
IC 1805 & 20 & 794 & 7--15 & 13.0& 0.042& 101\\
EO Per & 120 & 1138 & 8--17 & 14.5& 0.041 & 172\\

Gem OB1 & 4 & 456 & 6--14 & 11.5& 0.038& 43\\

\hline
\end{tabular}
\tablefoot{
Limiting $V$ magnitude (Lim.) is shown on Fig.~\ref{lim_mag} and represents the cutoff where stars have been detected on more than 90\% of the frames. The number of stars that have been detected >90\% (N$_{90}$) is the total number of good quality sources in each field.
}

\end{table}
\addtolength{\tabcolsep}{1pt}

\begin{figure*}[t]
        \includegraphics{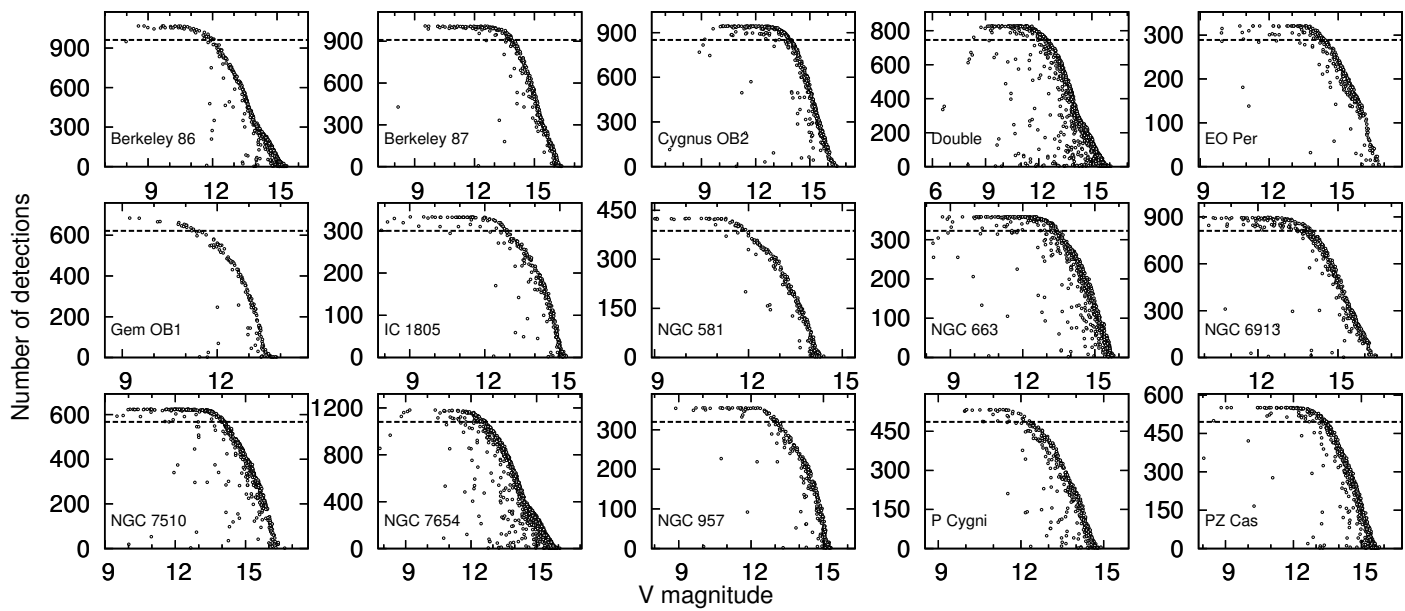}
        \caption{Number of detections per star as a function of $V$ magnitude for each of the fields. The dashed line represents the 90\% detection limit. We obtained the limiting $V$ magnitudes from the graphs by assessing the intersection of the 90\% line with the spine of the data. Stars above the 90\% line are considered as stars with good photometric quality.}
        \label{lim_mag}
\end{figure*}

The $V$ passband data are summarised in Table~\ref{field_stars}. 
We list the $V$ passband exposure times in column 2 (Exp.) and the total number of detected objects in column 3 (N).
The given $V$ magnitude range in column 4 is where the detected objects fall into.
The limiting $V$ magnitude (Lim.) values in column 5 are given as the limit where fainter stars are detected on less than 90\% of the frames taken for an individual observed field (cf. Fig.~\ref{lim_mag}).
Stars that are detected in over 90\% of the frames are defined as stars with good photometric quality (N$_{90}$).
These are stars for which the detection on our frames was not affected by the airmass or seeing values.
The total number of N$_{90}$ stars is around 2900.
The quick dropoff in detection for fainter stars is to be expected due to variable FWHM values and strict photometric detection limit.

Column 6 in Table~\ref{field_stars} lists the typical photometric internal root-mean-square (rms) noise ($\sigma$) of the constant stars to quantify our photometric precision at the limiting $V$ magnitude (Lim.).
The rms noise is defined by \citet{Kjeldsen:1992} and is calculated from the consecutive magnitudes measured during the same night.
We give the precision as an intersection on a $\sigma$-magnitude diagram between the internal noise and the limiting $V$ magnitude value.

\begin{figure}[t]
        \includegraphics{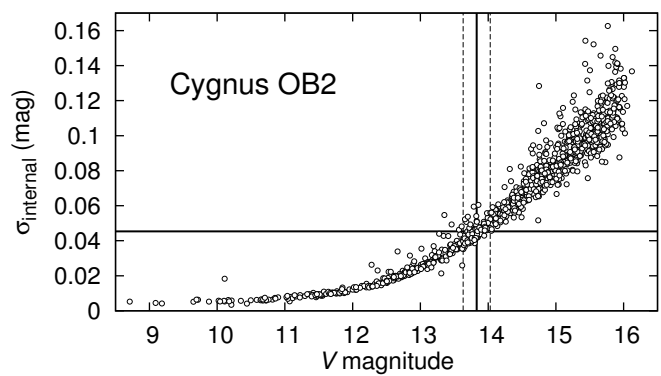}
        \caption{Internal rms noise as a function of magnitude in Cygnus OB2 field. The solid vertical line is the calculated limiting $V$ magnitude and the dashed vertical lines show the bin in which the median precision (horizontal line) is measured.}
        \label{scatter_mag}
\end{figure}

A representative $\sigma$-magnitude diagram of Cygnus OB2 field is shown in Fig.~\ref{scatter_mag}.
The resulting precision at limiting magnitude was obtained as a median internal noise from stars in a 0.4 magnitude bin around the limiting $V$ magnitude value (between the two dashed vertical lines).

\section{Detection of variability}

Variable stars were initially selected on the basis of a variability index.
We define the variability index as a ratio between the standard deviation of the full light curve and the internal rms noise \citep{Kjeldsen:1992}. 
For a constant star, the value would be close to unity whereas stars with values above 1.5 are counted as photometrically variable objects.
The variability index is useful for a wide variety of observations, as the internal rms noise takes into account all instrumental uncertainties and the resulting value is therefore independent of the total scatter of the data, effectively measuring only the signal-to-noise ratio of variability.

Throughout the variability determination, we use measurements in the $V$ passband because of the better signal-to-noise ratio for most of the sources. 
As the variability index is a measure of the signal-to-noise ratio of variability, we calculated the value of the variability index for most of our objects.
We only omitted stars that had fewer than 100 measurements from the selection to exclude very faint stars and spurious detections. 
For every star, we calculated the variability index based on three different data sets.
Two of these data sets were acquired by applying the two flat-field correction surfaces described in Section~\ref{sec:reduction} and the third was the raw light curve.
The final selection was based on the lowest variability index value of those data sets to minimise erroneous variability detections.

We then flagged our selected variable stars, based on the presence or absence of systematic effects, under different flags.
As the aforementioned problems requiring the flat-field correction were too severe in some cases, we had to flag a number of stars unusable when a clear separation between the data from the east and west side of the pier was seen.
Additional flagging was done based on the photometry extraction method used. 
We used aperture photometry for brighter sources and PSF photometry for fainter and crowded stars.

To analyse the light curve shapes of different type of variable stars, we used a Python library called Feature Analysis for Time Series (FATS).
It extracts up to 64 different features from an unequally sampled time series that are described in \citet{Nun:2015}.
We used some of these features as supplementary information to describe our data set.

\section{Results}

In total we found 354 variable stars in the observed fields.
We list objects in tables separated by variability type in Appendix \ref{sec:app_variables} with the full catalogue available at the CDS.

Table~\ref{tab:binaries_short} is an excerpt from the full table (Table~1 at CDS) of all the eclipsing binaries we found.
The table lists our field names and internal ID-s of the stars. 
This was added to help identify the stars in our finding charts that are added in the Appendix \ref{sec:app_field}.
The designations are based on the SIMBAD database and, in case the star was absent there, we used The Two Micron All-Sky Survey (2MASS) designations.
Given coordinates are based on 2MASS All-Sky Point Source Catalog\footnote{\url{http://irsa.ipac.caltech.edu/cgi-bin/Gator/nph-dd}}.

\begin{sidewaystable}
\caption{Observed eclipsing binaries.}
\label{tab:binaries_short}
\centering
\begin{tabular}{lclcccccccccc}
\hline 
\hline
\rule{0pt}{10pt}Field & ID & Designation & RA & DEC & Cluster & Spectral & $V$ & Amp. & Period & VSX & VSX per.\\
&  &  & J2000 & J2000 & membership & type & (mag) & (mag) & (days) & info  & (days)\\
\hline
\rule{0pt}{10pt}Berkeley 86 & 3 & V444 Cyg & 20:19:32.4 & +38:43:54 & -- & WN5+O6II-V & 7.973 & 0.336 & 4.2130 & EA/WR & 4.2124\\
Berkeley 86 & 22 & HD 228989 & 20:20:21.4 & +38:42:00 & -- & O8.5V:+O9.7V: & 9.686 & 0.091 & 1.7734 & -- & --\\
Berkeley 86 & 52 & LS II +38 43 & 20:19:36.8 & +38:35:21 & Berkeley 86 & B1V & 10.676 & 0.148 & 1.9102 & -- & --\\
Berkeley 86 & 289 & V435 Cyg & 20:16:27.0 & +38:45:41 & -- & B3V(e) & 12.749 & 0.659 & 6.7758 & EA/SD & 6.7719\\
Berkeley 86 & 381 & 2MASS J20190590+3853009 & 20:19:05.9 & +38:53:01 & -- & -- & 12.808 & 0.569 & 4.3262 & -- & --\\
Berkeley 87 & 18 & BD+36 4063 & 20:25:40.6 & +37:22:27 & -- & ON9.7Ib & 9.656 & 0.187 & 4.8121 & E/GS & 4.8126\\
Berkeley 87 & 50 & TYC 2697-130-1 & 20:25:30.6 & +37:25:48 & -- & F5V* & 10.736 & 0.169 & 0.7726 & EA & 0.7726\\
Berkeley 87 & 87 & V2538 Cyg & 20:23:33.5 & +37:25:45 & -- & B5:* & 12.003 & 0.496 & 3.1620 & EB & 3.1616\\
Berkeley 87 & 131 & TYC 2697-883-1 & 20:25:38.3 & +37:03:11 & -- & -- & 12.188 & 0.183 & 1.8632 & -- & --\\
Berkeley 87 & 203 & 2MASS J20222399+3658020 & 20:22:24.0 & +36:58:02 & Berkeley 87 & -- & 12.940 & 0.198 & 20.3009 & -- & --\\
\hline 
\end{tabular}
\tablefoot{
Full Table~\ref{tab:binaries_short} is available at the the CDS (Table~1). Similar tables are available for other variable stars. A portion of the table is given here to show the general style and form of the resulting data.
}
\end{sidewaystable}

The spectral types listed in Table~\ref{tab:binaries_short} are combined from SIMBAD database, the Tycho-2 spectral type catalogues \citep{Fabricius:2002,Wright:2003}, observations by \citet{Currie:2010}, and spectral classification compilations by \citet{skiff:2014} and \citet{Wright:2015}.
Spectral types with lower case letters are obtained from photometric observations in the Vilnius 7-colour system \citep{Mila:2013}.

\addtolength{\tabcolsep}{-2pt} 
\begin{table}
 \centering
\caption{Spectral classifications based on our observations.}
\label{sp_class}
\begin{tabular}{lcc}

\hline \hline
\rule{0pt}{10pt}Designation & Previous sp. type & Determined sp. type\\
  \hline 
\rule{0pt}{10pt}V1319 Cyg&--&K0I\\ 
TYC 2697-130-1&--&F5V\\
V2538 Cyg&--&B1III-V\\
TYC 2684-1267-1&--&K5V\\
J02163833+5705124&M2$^{(1)}$&M2:\\
BD+56 596&B$^{(2)}$&B0II-III\\
BD+56 733&OB+e:$^{(3)}$&B1II-III\\
TYC 3709-420-1&--&A1V\\
V472 Cas&--&M4:\\
BD+60 358&B5$^{(4)}$&B1Ve\\
\hline
\end{tabular}
\tablebib{
(1) \citet{Lee:1943};
(2) \citet{skiff:2014};
(3) \citet{Hardorp:1959};
(4) SIMBAD.
}

\end{table}
\addtolength{\tabcolsep}{2pt}

\begin{figure*}[t]
        \includegraphics[]{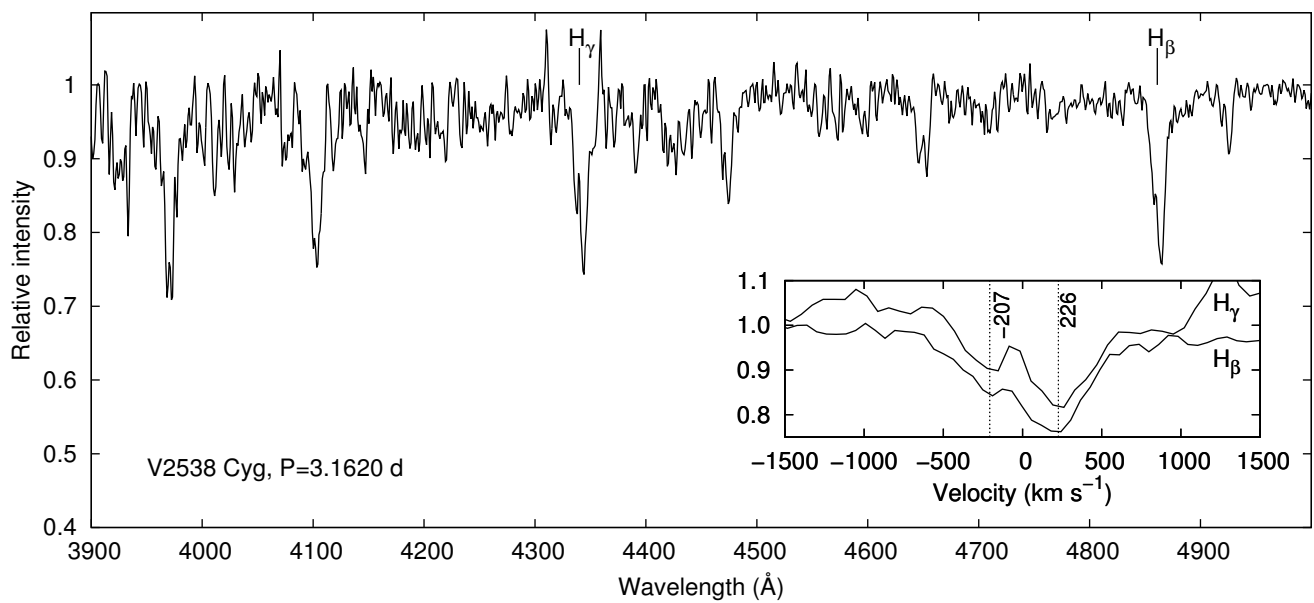}
        \caption{Spectrum of the double line B1III-V binary V2538 Cyg. The spectrum was obtained at phase 0.225 with 1.5 meter AZT-12 telescope at Tartu Observatory. Double lines are visible in all hydrogen and helium lines and the component velocities are shown in the inset panel.}
        \label{spectrum}
\end{figure*}

Based on our spectral observations from Tartu Observatory, we obtained spectral classifications for ten variable stars of which six had no previous classifications.
These classified variable stars are listed in Table~\ref{sp_class} and marked with an asterisk (*) behind the classification in Table~\ref{tab:binaries_short}.

From our spectra, BD+60 358 is a B1Ve type star as both H$\beta$ and Fe lines are in emission.
V472 Cas is an M4 or later type star and based on the Ca I 4227 line, its luminosity class is from V to III.
V2538 Cyg is a previously known eclipsing binary star \citep{Nicholson:2006} and the lines from its companion are also visible in our spectra, making it an SB2-type spectroscopic binary with a B1III-V type primary and a B-type secondary component (cf. Fig~\ref{spectrum}).
The spectra for V2538 Cyg was taken at a phase of 0.225 and the component radial velocities of -207 and 226 km s$^{-1}$ were estimated by Gaussian fitting the line components.

To calculate the cluster membership fraction among the variable stars, we used the MWSC project \citep{Kharchenko:2013} which incorporates cluster membership probabilities around open clusters.
They give three different probabilities based on proper motion and photometric $J$--$K$ and $J$--$H$ colour.
These probabilities were determined from the location of the stars with respect to the reference sequences of either the average cluster proper motion in kinematic diagrams or the isochrones in photometric diagrams.
They find about 400\,000 stars to be cluster members having kinematic and photometric membership probabilities higher than $60$\%.
We adopt their membership probabilities (kinematic probability (Pkin) $>50$\% and photometric probabilities (PJKs$+$PJH) $>100$\%) to determine the cluster membership of our variable stars.

In Table~\ref{tab:binaries_short} we compare our variables with previously known variability information on these objects. 
For this comparison we use the International Variable Star Index (VSX)\footnote{\url{https://www.aavso.org/vsx}} database run by the American Association of Variable Star Observers (AAVSO) to identify possible first discoveries of variability.
Due to the extensive compilation of different sources, the VSX database includes information on 398\,030 (as of July 2016) variable stars.
In the VSX "info" and "period" columns, we label stars that have no entry in the VSX database with "--" and those matches in the VSX catalogue with no variability classification with "?".
Table~\ref{tab:new_full} is a compilation of all the variable stars that are not listed in the VSX database.

In Table~\ref{tab:binaries_short}, we give an amplitude value as a modified FATS amplitude.
The amplitude value in the Table is defined as the difference between the median of the maximum 5\% and the median of the minimum 5\% magnitudes.

To search for periodic variables and eclipsing binaries from our data set, we used the discrete Fourier transform code SigSpec \citep{Reegen:2007}.
It calculates a significance value which describes the probability that an amplitude is not caused by noise.
SigSpec derives the resulting frequencies by iteratively prewhitening the frequency power spectrum using the most significant signal components from each run.
The dominant frequencies were confirmed by visually identifying an amplitude of at least twice of the folded phase curve scatter.

We used the significance of any given frequency to determine the probability of the signal being either from a pulsating star, an eclipsing binary or a false signal due to the aliasing from our window function.
We distinguish between a pulsational star and an eclipsing binary by visually inspecting the phase diagram.
In the case of suspected eclipsing binaries, we also used the phase dispersion minimisation method \citep{Stellingwerf:1978} to improve the period estimation as it accounts better for the different shapes of binary folded light curves than the Fourier method does.

In Table~\ref{tab:binaries_short}, we report only the periods corresponding to the dominant frequencies. We list
$\beta$ Cep and slowly pulsating B (SPB) stars with multiple frequencies in Table~\ref{multi_freq}.
A frequency power spectrum peak is commonly called significant when the signal-to-noise ratio of an amplitude is $S/N > 4$ \citep{Berger:1993} which corresponds to a significance value of 5.46 in SigSpec.
Following the more conservative approach by \citet{Chapellier:2011}, we list only the frequencies with significance value over 20 because the periodograms below this began to show a forest of peaks.

Ellipsoidal variables \citep{Morris:1985} are close binary stars that are non-spherical due to tidal interactions and exhibit sinusoidal phase curves due to the changes in the cross-sectional area that is visible to the observer.
In addition, some cases of overcontact binaries exhibit similar sinusoidal phase curves.
The phase curves of these binary stars are identical to pulsating variables and we only classified stars with light curves that show unequal minima or maxima as eclipsing binaries as this indicates tidal distortion effects or an uneven surface distribution of chemical elements \citep{Kourniotis:2014}. Variables with no identified periods and unknown spectral types were left unclassified.

\section{Classification of variable stars}

\addtolength{\tabcolsep}{-2pt} 
\begin{table}
 \centering
\caption{The observed variable stars by variability type.}
\label{variable_type}
\begin{tabular}{lcccc}

\hline \hline
\rule{0pt}{10pt}Variability & No. of stars & No. in & Cluster& No. of new\\
 type & observed & clusters & membership& discoveries\\
  \hline 
\rule{0pt}{10pt}EB& 80 & 18 & 23\% & 47\\
$\beta$ Cep& 13 & 3 & 23\% & 11\\
SPB& 16 & 3 & 19\% & 15\\ 
Be& 62 & 4 & 6\% & 29\\ 
$\alpha$ Cyg& 30 & 8 & 27\% & 9\\
WR& 3 & 0 & 0\% & 1\\ 
Cepheid& 7 & 0 & 0\% & 2\\ 
$\gamma$ Dor& 1 & 0 & 0\% & 0\\ 
SR & 63 & 8 & 13\% & 33\\ 
unclassified& 80 & 9 & 11\% & 74\\ 

\hline
\end{tabular}
\end{table}
\addtolength{\tabcolsep}{2pt}

Table~\ref{variable_type} summarises our observed variable stars.
The classification of variability was based on the period of variability as well as position on the HR diagram \citep{Aerts:2010}.
As we observed only the brighter objects in each field, the variability types we found are also from the upper part of the HR diagram, containing mostly $\alpha$ Cygni, $\beta$ Cephei, Be, SPB, Cepheid, semi-regular (SR) and Mira type variables.
The full tables listing all the members of each type are available in Appendix \ref{sec:app_variables}.

\subsection{Eclipsing binaries}

We classified eclipsing binaries based on their previous classifications from VSX or, in case of new discoveries, based on their non-sinusoidal phase curves.
In total, we found 80 eclipsing binary stars (Table~1 at CDS) in all of the fields, for 69 of which we identified their orbital periods.
The remaining 11 were named binary candidates as their light curves include some eclipse-like drops, but no period could be extracted due to the limited data.
According to VSX, 47 eclipsing binaries are new discoveries with no known period and variability designation.
Using our data set, we have previously analysed seven of the known eclipsing binaries in Cygnus OB2 association and found that A36, Schulte 5, $[$MT91$]$ 059, and $[$MT91$]$ 696 show a non-zero period change value at two sigma confidence \citep{Laur:2015}.

\subsection{$\beta$ Cephei stars}

\begin{table}
 \centering
\caption{$\beta$ Cephei and SPB stars with multiple frequencies.}
\label{multi_freq}
\begin{tabular}{llccc}

\hline \hline
\rule{0pt}{10pt}Type & Designation & Frequency & Amplitude& Sig.\\
&&$d^{-1}$&mag&\\
  \hline 
\rule{0pt}{10pt}$\beta$ Cep& LS II +36 70 & 4.63301 & 0.0164 & 89\\
&  & 4.77155 & 0.0074 & 44\\
&  & 4.52160 & 0.0059 & 30\\
&  & 4.14302 & 0.0059 & 22\\
\rule{0pt}{10pt}$\beta$ Cep& V665 Per & 4.12750 & 0.0241 & 51\\
&  & 3.98230 & 0.0117 & 42\\
\rule{0pt}{10pt}$\beta$ Cep& J23175230+ & 6.05643 & 0.0244 & 43\\
&+6025022  & 6.29459 & 0.0147 & 23\\
\rule{0pt}{10pt}$\beta$ Cep& LS III +60 58 & 4.67678 & 0.0141 & 67\\
&  & 4.51472 & 0.0052 & 20\\

\rule{0pt}{10pt}SPB& HD 229253 & 0.636170 &0.0050 & 36 \\
&  & 0.289096 & 0.0042 & 26\\
\rule{0pt}{10pt}SPB& TYC 2684-312-1 & 1.16245 & 0.0130 & 41 \\
&  & 0.812356 & 0.0108 & 28\\
\rule{0pt}{10pt}SPB& [MT91] 626 & 0.900873 & 0.0174 & 36\\
&  & 0.905040 & 0.0152 & 35\\

\hline
\end{tabular}
\tablefoot{
Sig. - Significance from SigSpec
}
\end{table}

$\beta$ Cephei stars are pulsating early-to-mid B-type dwarfs to giants with periods ranging between two and eight hours.
The total number of known $\beta$ Cephei stars is around 200 \citep{Pigulski:2008}.

We find 13 $\beta$ Cephei candidates (Table~2 at CDS) of which two have been previously listed in the VSX database and eight have a known spectral type.
We report four $\beta$ Cephei stars that have multiple frequencies over the significance value of 20 (Table~\ref{multi_freq}).

The only $\beta$ Cephei star in our selection with a previously determined period is V665 Per.
\citet{Gomez-Forrellad:2000} reported two periods of 0.19493 and 0.24233 days.
Their first period can be found in our data with significance value of eight.
Their second period is closest to our dominant period of 0.242277~d but we would like to point out that our data do not support their period of 0.24233~d.
This difference in periods can be interpreted by a period decrease of around five seconds in 15 years ($\dot{P}/P=-0.24$ $\mathrm{Myr^{-1}}$).
This period change rate could be used to constrain evolutionary models for $\beta$ Cephei stars by assuming different amounts of initial rotation and convective core overshoot and comparing the theoretical rate of period change to the measured one as done in \citet{Neilson:2015}.

\subsection{SPB stars}

SPBs consist of mid-to-late B-type dwarf to giant stars with periods roughly between 0.8 and 3 days \citep{Aerts:2010}.

We find 16 possible SPBs (Table~3 at CDS) based on their periods, of which 12 also have a spectral classification in the literature.
We report three SPB stars with two frequencies over significance value of 20 (Table~\ref{multi_freq}).

\begin{figure*}[t]
        \includegraphics[]{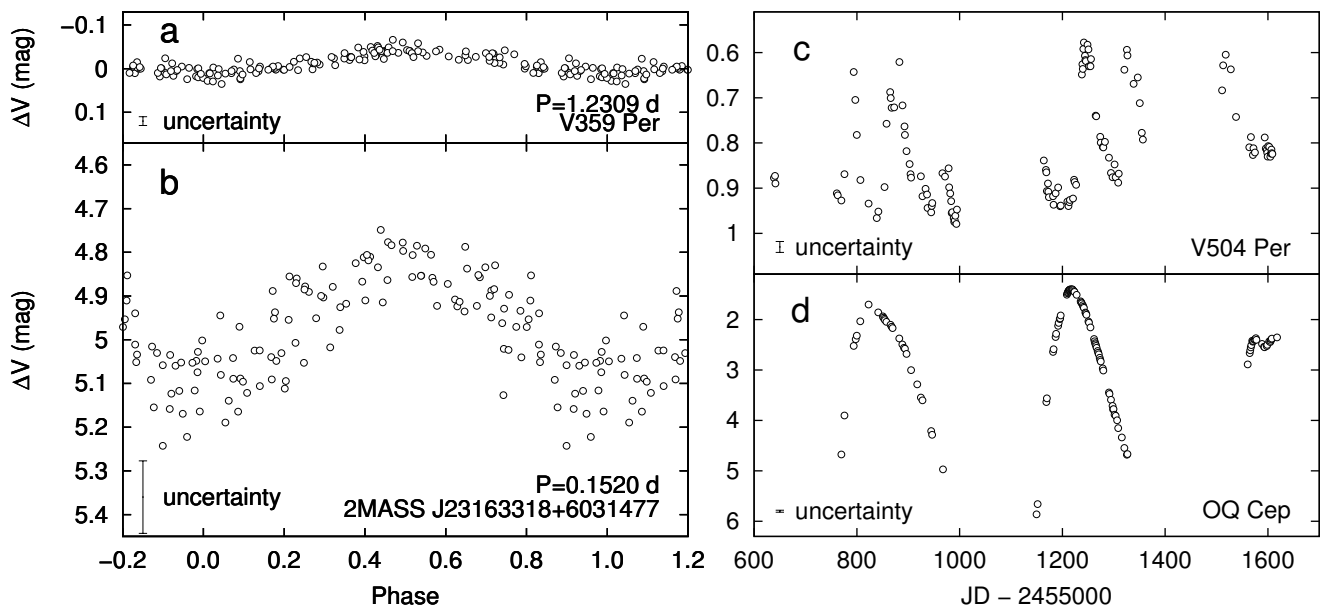}
        \caption{Example light curves of four variable stars from nightly mean magnitudes. The leftmost light curves are phased SPB (panel a) and $\beta$ Cep type (panel b) with the latter being the faintest periodic star in our survey. The rightmost light curves are Be (panel c) and SRb type (panel d).}
        \label{lightcurve_example}
\end{figure*}

Our only matching SPB in the VSX database, V359 Per, is listed there as a $\beta$ Cep star.
The previous determination is done by the IOMC pipeline \citep{Alfonso:2012} and they find a period of 0.275~d\footnote{\url{http://sdc.cab.inta-csic.es/omc/var/3694000069.html}}.
This period is not supported by visual inspection of the IOMC phase curve nor by our data.
We find a different period of 1.2309~d, which is more in line with the star being an SPB type (cf. Fig.~\ref{lightcurve_example}).

\subsection{Be stars}

While some Be stars show oscillations that are in accordance with either $\beta$ Cep or SPB stars \citep{Walker:2005,Balona:1995}, Be stars infrequently exhibit photometric outburst behaviour linked to the dispersion of the circumstellar disk \citep{Rivinius:2013}.

As Be stars are classified from their spectra, we have compiled here every variable B star that has shown some kind of emission in the past. 
In total, we find 62 Be stars (Table~8 at CDS) in our selection to be variable.
Two of them, TYC 4279-2067-1 and V356 Per, exhibit additional $\beta$ Cep-like pulsations with periods of 0.582~d and 0.342~d, respectively.
V356 Per is previously classified by IOMC pipeline as a $\beta$ Cep candidate\footnote{\url{http://sdc.cab.inta-csic.es/omc/var/3694000050.html}}.
They also find a period of 0.30595~d which is not supported by our data.
Instead, we find a period of 0.342~d with a phase curve that is more sinusoidal than the IOMC data.

Be stars with photometric outbursts are indicated as "Y" in the outburst column in Table~8 at CDS.
Outbursts can be caused by a heating front that propagates across the disk and, after reaching the critical density, pushes the disk into a hot state with high mass transfer and high accretion luminosity.
\citep{Mennickent:2002} distinguish four types of photometric behaviour with type 1 being caused by accretion of matter with lower mass transfer rates and type 2 by a sustained period of a high mass-transfer rate.
27 of our Be stars exhibit outbursts with sudden rises and declines (type 1) during the three seasons.
The highest amplitude outburst of 0.87 mag was seen in EO Per during the 2012 season.
In addition, VES 203 exhibits behaviour of sudden brightness jumps (type 2) where the brightness dropped 0.2 mag for 100 days in the beginning of 2012 season and then stayed at a lower level for the remainder of the campaign.

\subsection{$\alpha$ Cygni stars}

$\alpha$ Cygni variables are non-radially pulsating supergiants of B--A spectral types with irregular variability caused by the superposition of many oscillations with close periods \citep{Aerts:2010,Saio:2013}.
The typical observed timescales range from 1 to 50 days \citep{Eyer:2008}.

We classified all of our variable B-to-early-A type supergiants (luminosity class I and II) as $\alpha$ Cygni variables.
In total, we find 30 $\alpha$ Cygni variables (Table~7 at CDS) and for three of them, HD 229059, HD 14143 and V757 Per, we find pulsation periods.
HD 14143 and V757 Per have previous period determinations of 0.0892 \citep{Koen:2002} and 0.1643 \citep{DeCat:2007} days, accordingly, although we found different period values with 0.5321 and 0.2453 days.
Altogether, VSX lists periods for six of our determined $\alpha$ Cygni variables although none of those periods can be seen in our data.
We may not have observed the same periods in our data because the variability of $\alpha$ Cygni type stars may be cyclic in nature, caused by interior mixing processes and atmospheric instabilities \citep{Aerts:2010}.

\subsection{Wolf-Rayet stars}

Almost all Wolf-Rayet (WR) stars show no strict periodicities and the origin of their variability remains poorly understood \citep{Aerts:2010}.

According to the spectral classification, we find three variable WR stars (Table~6 at CDS): WR 141, WR 158 and V444 Cyg.
WR 141 and WR 158 exhibit irregular variability with amplitudes of 0.108 and 0.073 mag, respectively.
WR 158 is previously known spectroscopically variable star \citep{St-Louis:2009}.
WR 141 is previously classified as an SB2 binary \citep{Marchenko:1998} with an O5V--III companion and an orbital period of 21.698~d.
Only from the first season of our observations do we find some signal with a period of 21.7~d, whereas there is no signal from the other seasons.
This might be due to the flickering nature of WR 141 \citep{Khalack:1996}.
V444 Cyg is a previously known eclipsing binary \citep{Cherepashchuk:1973} with a WR component.
Its orbital period from IOMC data is 4.213~d\footnote{\url{http://sdc.cab.inta-csic.es/omc/var/3151000080.html}} which can be confirmed with our data.

\subsection{Late-type stars}

Variable stars situated on the red side of the classical instability strip and with long periods (P > 80 days) are classified as Mira or SRa type variables if they are periodic, and SRb type if they possess alternating intervals of periodic and slow irregular changes in their light curves \citep{Aerts:2010,Eyer:2012}. 
F--K spectral type weak-lined giants and supergiants are classified as SRd type variables.
As we monitored our fields for three seasons our time series are not long enough to comfortably determine very long periods and therefore we classified all of the late-type stars into one group.

There are 46 variable stars in our fields with late-type spectral classification.
Based on our measured $V$--$I$ colours and the shape of the light curves, we find additional 17 late-type candidates (Table~9 at CDS) with $V$--$I$ over 1.8 based on the effective temperature and colour tables from \citet{Currie:2010}. 
Due to the varying degree of reddening in young clusters, some of the candidates may be included because of the reddening difference between the comparison stars and target stars.
We tried to account for that by calculating the differential reddening of the comparison stars by colour transforming the known APASS and 2MASS colours in the field.

From the total of 63 late-type variables, 30 are listed in the VSX database with 16 of them having a determined period.
We find semi-regular periods for 19 stars only three of which are close to the previously determined periods.
Some periods were obtained after de-trending the light curve with a wide median-filter to smooth out the slow irregular changes in semi-regular stars.
As the periods are semi-regular and vary in time, the discrepancy between our newly determined and the previously determined periods is understandable.

\subsection{Cepheids}
Cepheids are radially pulsating F--K type giants and supergiants and their luminosity classes change roughly from III to I during the pulsation cycle \citep{Aerts:2010,Turner:2012}.
A typical light curve of a Cepheid is skewed and periodic, ranging from 1 to 50 days.

We find seven Cepheid variables (Table~4 at CDS) of which five were previously known.
V1319 Cyg has been previously classified as an SRa type, probably due to its longer period.
On the other hand, the folded light curve shows a classical skewed Cepheid shape and the spectral classification of K0I, determined from our observations at Tartu Observatory, is in accordance with a Cepheid rather than SR type of star.
Of the two previously unknown Cepheid candidates, NGC 6913-36 exhibits classical Cepheid phase curve shape and has a spectral class estimation of "f" (photometrically determined F type) based on Vilnius photometric system \citep{Mila:2013}.
The second candidate, 2MASS J20173903+3804143, is added as a candidate because of its brightness, period length and phase curve shape, although the photometric scatter is too high to distinguish between a noisy sinusoidal shape and a classical Cepheid shape.

\begin{figure}[t]
        \includegraphics{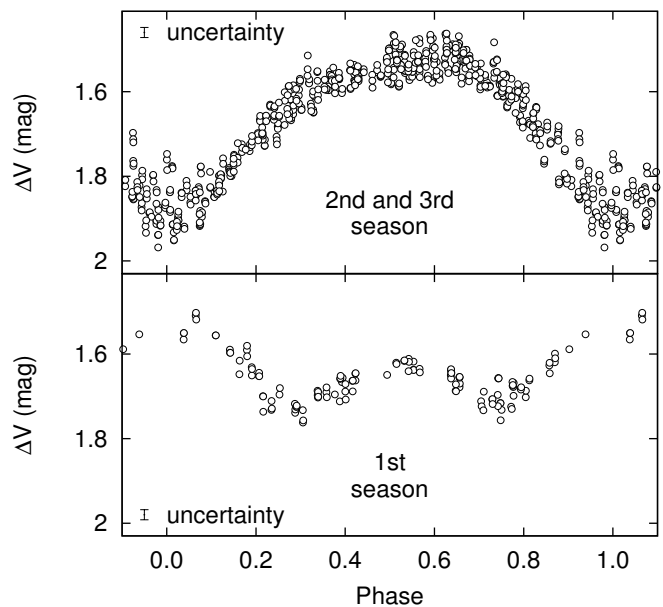}
        \caption{Folded light curve of a Cepheid NSVS 5731720. The data in the lower panel are from the first season and in the upper panel from second and third season. Both of them have been folded with the same period of 3.20612. The first season shows binary-like shape and a phase shift of around 0.6.}
        \label{cep_example}
\end{figure}

We note that for the Cepheid NSVS 5731720, only the last two seasons can be used to fold the light curve nicely with the period (Fig.~\ref{cep_example}).
The first season has approximately half the amplitude of the latter two, and when folded with the same period of 3.20612 days, the $V$ passband light curve exhibits a binary-like shape with two unequal maxima.
There is also a phase shift of 0.6 between the first two seasons. 
We are unable to explain this behaviour with systematic effects.

\subsection{Other variables}

There is one previously known $\gamma$ Doradus star V1132 Cas (Table~5 at CDS) for which we confirm the period of 0.626 days \citep{Wyrzykowski:2002}.
As $\gamma$ Dor stars have masses of $1.5-1.8 M_{\odot}$ and are of luminosity type IV--V \citep{Aerts:2010}, the apparent magnitude of 8.16 mag for V1132 Cas suggests that it is a nearby field star.

There are an additional 80 variable stars that we could not classify due to the lack of spectral or pulsational information (Table~10 at CDS).
V1156 Cas and TYC 3695-992-1 have previously determined periods but as the stars are too faint in our data we were unable to confirm them.
We find periods for 13 of these unclassified objects (Table~10 at CDS), ranging from 0.5 to 102 days.

\subsection{Examples}

We present some light curve examples in Fig.~\ref{lightcurve_example}.
Phase curves in panels a and b show the difference in scatter for one bright and one faint star for which we managed to find a period.
V359 Per is a SPB star with $V$ magnitude of 9.06 and shows a scatter of around 0.02 mag.
2MASS J23163318+6031477 is a fainter $\beta$ Cep star with $V$ magnitude of 15.188 and shows a scatter of around 0.15 mag.
While it is below our limiting magnitude for the NGC 7510 field, the high amplitude of 0.489 mag is enough to detect the period.
We did not find any previous variability information on this star from any database, probably due to its low brightness

Panels c and d in Fig.~\ref{lightcurve_example} are examples of irregular or semi-regular behaviour.
V504 Per is a Be star with one of the most numerous outburst counts, showing five major outbursts during the observing campaign.
OQ Cep is an M7 type star exhibiting SR variation of over four magnitudes during the second observing season, the highest amplitude variation across all our data.
We also find a semi-regular period of 385 days, based on the first two seasons, which is comparable to the previous period determination of 370 days \citep{Alfonso:2012}.

\begin{figure}[t]
        \includegraphics[width=0.49\textwidth, trim=0 0 0 0,clip]{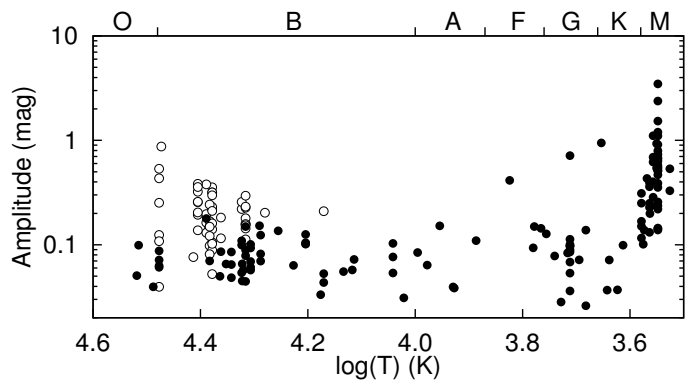}
        \caption{Light curve amplitude as a function of effective temperature for all variables except eclipsing binaries. Be stars are marked with open circles and others with filled circles.}
        \label{logT_amp}
\end{figure}

We present the amplitude distribution of variable stars in Fig.~\ref{logT_amp}.
The effective temperatures corresponding to specific spectral types found in the literature are acquired from Tables 7--9 in \citet{Currie:2010}.
In case there was no luminosity class determined, their temperature is given as their spectral subgroup mean temperature.
The figure shows that the late-type stars have a larger variability amplitude compared to the rest of the selection.
Be stars skew the amplitude of hotter stars towards higher values due to their outbursts but as their variability is caused by the circumstellar disk instead of their intrinsic pulsations, they are marked separately with open circles.

\section{Discussion}

To test the suitability of the matches to eclipsing binary and pulsating variable categories, we use the FATS skewness and kurtosis statistics.
Eclipses in eclipsing binaries cause the magnitude distribution to have a tail towards the fainter magnitudes and therefore positive skewness values, whereas pulsating stars have a close to zero or negative skewness values due to a more even magnitude distribution.
This separates the two groups of periodic variability, eclipsing binaries and pulsating stars, on the skewness-kurtosis plane.

\begin{figure}[t]
        \includegraphics[width=0.49\textwidth, trim=0 0 0 0,clip]{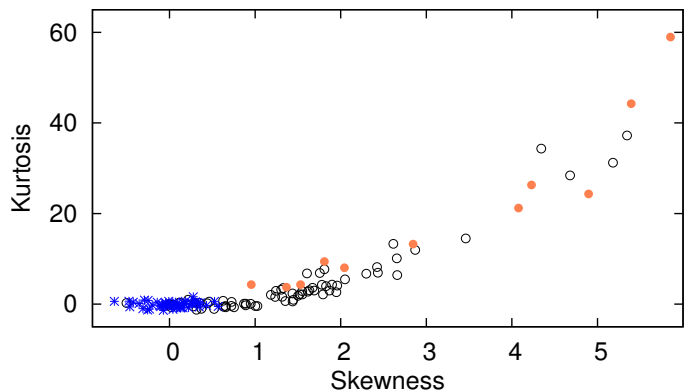}
        \caption{Skewness-Kurtosis diagram for the eclipsing binaries (circles) and pulsating variables (star icons). Binary candidates are filled circles, following the skewness-kurtosis trend of other binaries.}
        \label{skewness-kurtosis}
\end{figure}

Fig.~\ref{skewness-kurtosis} shows the skewness-kurtosis plane for the eclipsing binaries and pulsating stars we detected in our fields. 
The pulsating stars clump up on the left hand of the diagram and the eclipsing binaries are in a long tail in the right-upper part of the diagram, with the two types having a separation around a skewness value of $0.5$ as indicated by \citet{Graczyk:2010}.
The collection of binary stars up to a skewness value of 1 consists of ellipsoidal variables that show sinusoidal light curves similar to other pulsating stars but are a little more skewed due to irregularities in the uneven maxima of two consecutive periods.
The coloured circles represent our binary candidates (where no orbital periods were found) and they follow the skewness-kurtosis trend of the other binaries.

Despite finding variability in previously known objects, our photometric depth does not always allow us to find periodicities to compare with already published values.
\citet{Saesen:2013} monitored NGC 884 over three observing seasons and found periods for 131 stars in their field.
Eight of them match our listed variables and of these, we determined a period for only one object, due to the limiting magnitude in the field of our data.

\begin{figure*}[t]
        \includegraphics[]{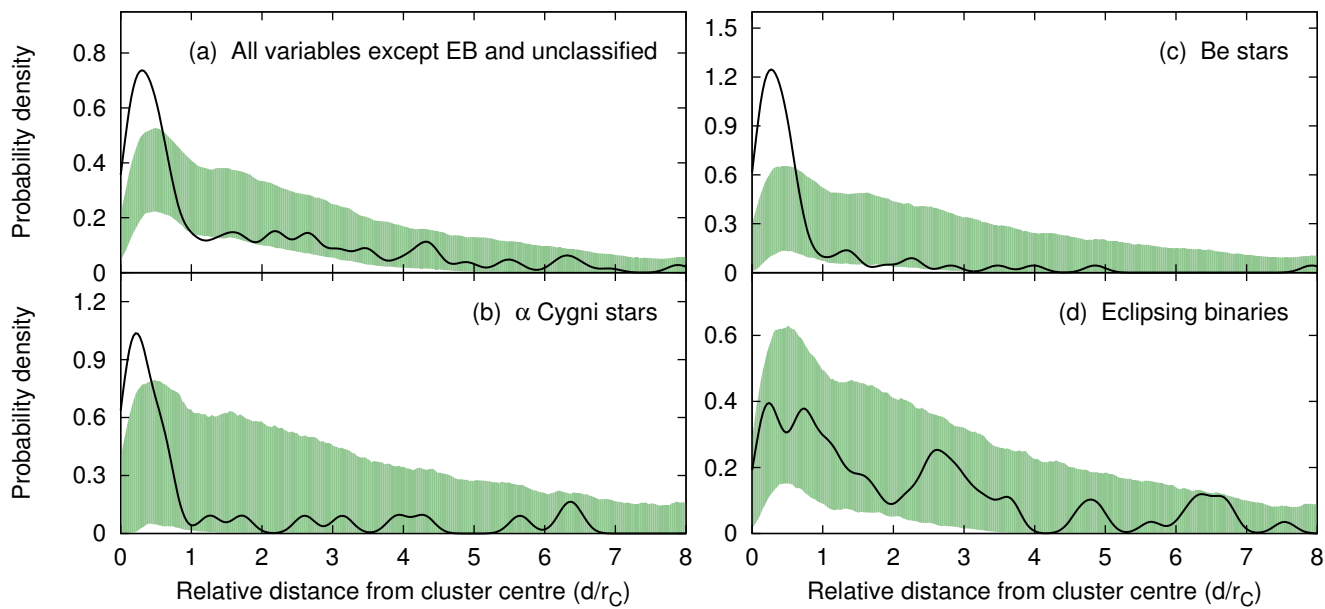}
        \caption{Probability density of variable stars at relative distance from the cluster centre. The coloured area represents the 99.7\% confidence limits calculated from all stars in our fields. The solid line is the probability density calculated only from the labelled types of variables.}
        \label{pdf}
\end{figure*}

Omitting eclipsing binaries, 35 out of the 275 intrinsically variable stars were found to be probable cluster members, based on the cluster membership probabilities of individual objects from Milky Way Star Clusters Catalogue (MWSC) \citep{Kharchenko:2013}.
This makes the total cluster membership fraction for variable stars less than 13\% (membership fractions for individual types are given in Table~\ref{variable_type}).
Using a limiting magnitude of 14 mag, that is, close to the limiting magnitude of our observations, we find a total of 21,792 stars around our observed clusters in the MWSC database.
Applying the same membership criteria as we did with our variable stars (kinematic probability (Pkin) $>50$\% and photometric probabilities (PJKs$+$PJH) $>100$\%), we find that 11\% of all those stars are cluster members. 
Thus the cluster membership fraction among variable stars is comparable to all of the stars around clusters according to the MWSC.

One of our goals in monitoring stellar clusters of different ages was to study massive stars in various evolutionary phases.
To find independent stellar ages, we need accurate membership information for these stars.
To this end, we compared MWSC catalogue to other works to test the reliability of their cluster membership estimations.
\citet{Wright:2015} studied 167 probable cluster members in the Cygnus OB2 association.
Matching their 167 cluster members with MWSC database, we find that out of 155 matched stars, only 51 are probable cluster members according to MWSC.

The DAML02 database \citep{Dias:2002} also includes membership probabilities for stars around clusters although their sample size is much smaller than in MWSC.
For example, in Berkeley 87 open cluster DAML02 has 157 nearby stars whereas MWSC has 23567.
Matching the two data sets, we find 141 common stars.
For this sample, DAML02 gives a membership fraction of 72\% (membership probability over 50\%) whereas MWSC gives a fraction of 29\%. This shows that MWSC underestimates the cluster membership fraction compared to other works.
But as it is the only database that covers most of our observed stars, we were obliged to use it for its completeness.

In order to find an upper limit to the cluster membership fraction, we studied the spatial distribution of variable stars in the plane of the sky and measured how variable stars are positioned around stellar clusters.
For each star, we calculated the relative distance to the centre of its closest cluster $d'=d/r_{C}$, where $d$ is the angular distance and $r_{C}$ is the angular cluster radius from MWSC.
This cluster radius is defined as the distance from the cluster centre to where the surface density of stars becomes equal to the average density of the surrounding field.
Based on these distances, we calculated the probability densities of finding a star at a certain relative distance from the centre of any cluster.
We used the bootstrapping technique to get the probability density of all stars with 99.7\% (3$\sigma$) confidence limits.

The resulting probability density functions (PDF) can be seen in Fig.~\ref{pdf}.
Panel (a) represents the combined distribution of $\alpha$ Cyg, $\beta$ Cep, Be, SPB, Cepheid, $\gamma$ Doradus, Wolf-Rayet and late-type variables.
This distribution shows that there are relatively more variable stars inside one unit of cluster radius than there are non-variable stars (represented by the coloured area).
Panels b and c show the PDF for $\alpha$ Cygni and Be type stars, respectively.
They both display the same signal as the combined distribution, albeit the 3$\sigma$ confidence area is larger due to smaller sample size.
The PDF on panel d, on the other hand, consists of eclipsing binaries and shows no deviation from the 3$\sigma$ level.
This means that eclipsing binaries have no positional preferences in relation to the clusters.
As the variability of binaries is not intrinsic, we omitted them together with unclassified variables from panel (a).

The distribution of all stars shows a maximum at around 0.5 $r_{C}$ with the width of the excess around 1 $r_{C}$ which is to be expected from the definition of the cluster radius.
The maximum of our observed variable stars in the plane of the sky is at around 0.3 $r_{C}$, meaning that the stars in the core of the clusters are more likely variable than the stars at the edges of the clusters.

Integrating the PDF of variable stars to one cluster radius, we find that 35\% of variable stars are located inside the clusters.
As we studied the spatial distribution in the plane of the sky, this can only be the upper limit of the cluster membership fraction due to fore- and background object contamination.

These examples indicate that the cluster membership probabilities in MWSC might be too strict and thus the membership fraction of 13\% amongst our variable stars is underestimated and can go as high as 35\%.
We did not use DAML02 database for cluster memberships, as their sample size is much smaller and did not include many of our variable stars.
But luckily, the reliability of membership probabilities will improve with the arrival of Gaia's more accurate distance, proper motion and radial velocity data as it measured even the brightest of sources in our fields \citep{gaia:2016}.

\section{Conclusions}

We monitored 22 northern open clusters and associations in 2011--2013 across 15 fields. 
Using a small-aperture telescope, we optimised our observations for the brightest stars in each field and obtained light curves for $\sim$3000 stars which were detected on more than 90\% of the frames.
We used a discrete Fourier transform code called SigSpec to obtain periods for pulsating stars and phase dispersion minimum algorithm to fine-tune the periods for eclipsing binaries.

Based on the variability index method we found 354 variable stars in the fields.
In addition to the pulsational information, we used a pulsation HR diagram to classify our variable stars. 
We classified 80 eclipsing binaries, 31 $\alpha$ Cyg, 13 $\beta$ Cep, 62 Be, 16 SPB, 7 Cepheid, 1 $\gamma$ Doradus, 3 WR and 63 late-type stars.
We compared our results with the Variable Star Index (VSX) to identify possible first discoveries of variability amongst the detected variable stars.
We found 47 eclipsing binaries, 10 $\alpha$ Cyg, 11 $\beta$ Cep, 29 Be, 15 SPB, 2 Cepheid, 1 WR and 33 late-type stars that are absent in the VSX database.
An additional 70 variable stars could not be classified due to the lack of spectral or pulsational information.

We find that 13\% of the variable stars are cluster members based on the membership information in the Milky Way Star Clusters Catalogue (MWSC).
The MWSC membership estimations are lower compared to other membership estimations, but we had to use it as it is the only database that covers most of our observed stars.
We analysed the probability density function of variable stars and found the upper limit of membership fraction to be 35\%. We list all 354 variable stars in CDS divided into different variability types.
In addition, the light curve data will also be released in the CDS.

\begin{acknowledgements}
This work was supported by institutional research funding of the Estonian Ministry of Education and Research (IUT26-2, IUT40-1, IUT40-2), by the Estonian Research Council grants 8906 and PUTJD5, and by the European Regional Development Fund (TK133).
This research has made use of the International Variable Star Index (VSX) database, operated at AAVSO, Cambridge, Massachusetts, USA.
Figures were made using the Gnuplot plotting utility and APLpy, an open-source plotting package for Python hosted at http://aplpy.github.com. 
The use of SIMBAD database and AAVSO Photometric All-Sky Survey (APASS), funded by the Robert Martin Ayers Sciences Fund. are acknowledged.
We thank an anonymous referee for constructive suggestions that have helped to improve our paper.

\end{acknowledgements}

\bibliographystyle{aa} 
\bibliography{Laur2016} 
\begin{appendix}

\section{Variable star catalogue}\label{sec:app_variables}

The full variable star catalogue is available at the CDS along with light curves.

\onecolumn
\addtolength{\tabcolsep}{-1pt} 
\begin{landscape}
\begin{longtable}{lclcccccccccc}
\caption{Observed eclipsing binaries.}\\
\label{tab:binaries_full}\\
\hline 
\hline
\rule{0pt}{10pt}Field & ID & Designation & RA & DEC & Cluster & Spectral & $V$ & Amplitude & Period & VSX & VSX period\\
&  &  & J2000 & J2000 & membership & type & (mag) & (mag) & (days) & info  & (days)\\
\hline
\endfirsthead
\caption{continued.}\\
\hline \hline
\rule{0pt}{10pt}Field & ID & Designation & RA & DEC & Cluster & Spectral & $V$ & Amplitude & Period & VSX & VSX period\\
&  &  & J2000 & J2000 & membership & type & (mag) & (mag) & (days) & info  & (days)\\
\hline
\endhead
\hline 
\endfoot
\rule{0pt}{10pt}Berkeley 86 & 3 & V444 Cyg & 20:19:32.4 & +38:43:54 & -- & WN5+O6II-V & 7.973 & 0.336 & 4.2130 & EA/WR & 4.2124\\
Berkeley 86 & 22 & HD 228989 & 20:20:21.4 & +38:42:00 & -- & O8.5V:+O9.7V: & 9.686 & 0.091 & 1.7734 & -- & --\\
Berkeley 86 & 52 & LS II +38 43 & 20:19:36.8 & +38:35:21 & Berkeley 86 & B1V & 10.676 & 0.148 & 1.9102 & -- & --\\
Berkeley 86 & 289 & V435 Cyg & 20:16:27.0 & +38:45:41 & -- & B3V(e) & 12.749 & 0.659 & 6.7758 & EA/SD & 6.7719\\
Berkeley 86 & 381 & 2MASS J20190590+3853009 & 20:19:05.9 & +38:53:01 & -- & -- & 12.808 & 0.569 & 4.3262 & -- & --\\
Berkeley 87 & 18 & BD+36 4063 & 20:25:40.6 & +37:22:27 & -- & ON9.7Ib & 9.656 & 0.187 & 4.8121 & E/GS & 4.8126\\
Berkeley 87 & 50 & TYC 2697-130-1 & 20:25:30.6 & +37:25:48 & -- & F5V* & 10.736 & 0.169 & 0.7726 & EA & 0.7726\\
Berkeley 87 & 87 & V2538 Cyg & 20:23:33.5 & +37:25:45 & -- & B1III-V* & 12.003 & 0.496 & 3.1620 & EB & 3.1616\\
\end{longtable}
\addtolength{\tabcolsep}{1pt}

\begin{longtable}{lclcccccccccc}
\caption{Observed $\beta$ Cephei variables.}\\
\label{tab:BCEP_full}\\
\hline 
\hline
\rule{0pt}{10pt}Field & ID & Designation & RA & DEC & Cluster & Spectral & $V$ & Amplitude & Period & VSX & VSX period\\
&  &  & J2000 & J2000 & membership & type & (mag) & (mag) & (days) & info  & (days)\\
\hline
\endfirsthead
\caption{continued.}\\
\hline \hline
\rule{0pt}{10pt}Field & ID & Designation & RA & DEC & Cluster & Spectral & $V$ & Amplitude & Period & VSX & VSX period\\
&  &  & J2000 & J2000 & membership & type & (mag) & (mag) & (days) & info  & (days)\\
\hline
\endhead
\hline 
\endfoot
\rule{0pt}{10pt}Berkeley 86 & 5 & HD 193032 & 20:16:40.4 & +38:53:57 & -- & B0.2III & 8.308 & 0.071 & 0.1808 & ? & --\\
Berkeley 86 & 28 & LS II +38 39 & 20:18:59.4 & +38:57:14 & -- & B1III & 10.057 & 0.179 & 0.2286 & -- & --\\
Berkeley 86 & 323 & 2MASS J20194609+3851325 & 20:19:46.1 & +38:51:32 & -- & -- & 12.855 & 0.372 & 0.1381 & -- & --\\
Berkeley 87 & 55 & LS II +36 70 & 20:22:53.3 & +37:06:46 & -- & OB- & 11.093 & 0.063 & 0.2158 & -- & --\\
Berkeley 87 & 338 & UCAC3 255-208501 & 20:24:26.8 & +37:19:33 & -- & -- & 13.468 & 0.333 & 0.1986 & -- & --\\
NGC 6913 & 17 & HD 229171 & 20:23:02.9 & +38:27:21 & -- & B2IV & 9.380 & 0.057 & 0.2855 & -- & --\\
NGC 7510 & 40 & LS III +60 58 & 23:16:12.3 & +60:27:24 & Markarian 50 & OB & 10.663 & 0.053 & 0.2138 & -- & --\\
NGC 7510 & 84 & LS III +60 59 & 23:16:33.1 & +60:28:48 & -- & OB & 11.629 & 0.050 & 0.1928 & -- & --\\
\end{longtable}

\begin{longtable}{lclcccccccccc}
\caption{Observed SPB variables.}\\
\label{tab:SPB_full}\\
\hline 
\hline
\rule{0pt}{10pt}Field & ID & Designation & RA & DEC & Cluster & Spectral & $V$ & Amplitude & Period & VSX & VSX period\\
&  &  & J2000 & J2000 & membership & type & (mag) & (mag) & (days) & info  & (days)\\
\hline
\endfirsthead
\caption{continued.}\\
\hline \hline
\rule{0pt}{10pt}Field & ID & Designation & RA & DEC & Cluster & Spectral & $V$ & Amplitude & Period & VSX & VSX period\\
&  &  & J2000 & J2000 & membership & type & (mag) & (mag) & (days) & info  & (days)\\
\hline
\endhead
\hline 
\endfoot
Berkeley 87 & 118 & TYC 2684-312-1 & 20:23:24.7 & +37:13:54 & -- & -- & 11.973 & 0.084 & 0.8602 & -- & --\\
Berkeley 87 & 240 & 2MASS J20234453+3713410 & 20:23:44.5 & +37:13:41 & Berkeley 87 & -- & 13.048 & 0.146 & 3.0662 & -- & --\\
Cygnus OB2 & 81 & $[$MT91$]$ 615 & 20:33:42.1 & +41:07:54 & -- & B2V & 11.405 & 0.079 & 0.8139 & -- & --\\
Cygnus OB2 & 190 & $[$MT91$]$ 626 & 20:33:45.0 & +41:22:32 & -- & -- & 12.541 & 0.108 & 1.1100 & -- & --\\
Cygnus OB2 & 229 & $[$MT91$]$ 29 & 20:30:53.5 & +41:27:01 & -- & -- & 12.813 & 0.120 & 1.3136 & -- & --\\
Gemini OB1 & 44 & HD 252944 & 06:11:27.7 & +23:13:53 & -- & B9 & 10.869 & 0.228 & 0.3852 & -- & --\\
NGC 6913 & 30 & HD 229253 & 20:24:25.2 & +38:29:07 & -- & B5IV & 10.155 & 0.033 & 1.5719 & -- & --\\
NGC 6913 & 36 & LS II +38 67 & 20:22:41.8 & +38:10:38 & NGC 6913 & B3III & 10.642 & 0.064 & 0.7952 & -- & --\\
\end{longtable}

\begin{longtable}{lclcccccccccc}
\caption{Observed Cepheid variables.}\\
\label{tab:CEP_full}\\
\hline 
\hline
\rule{0pt}{10pt}Field & ID & Designation & RA & DEC & Cluster & Spectral & $V$ & Amplitude & Period & VSX & VSX period\\
&  &  & J2000 & J2000 & membership & type & (mag) & (mag) & (days) & info  & (days)\\
\hline
\endfirsthead
\caption{continued.}\\
\hline \hline
\rule{0pt}{10pt}Field & ID & Designation & RA & DEC & Cluster & Spectral & $V$ & Amplitude & Period & VSX & VSX period\\
&  &  & J2000 & J2000 & membership & type & (mag) & (mag) & (days) & info  & (days)\\
\hline
\endhead
\hline 
\endfoot
\rule{0pt}{10pt}Berkeley 86 & 65 & V1319 Cyg & 20:20:51.1 & +39:00:45 & -- & K0I* & 11.240 & 0.944 & 41.3353 & SRA & 41.8700\\
NGC 663 & 28 & BY Cas & 01:47:11.9 & +61:25:21 & -- & F4Ib & 10.369 & 0.414 & 3.2221 & DCEPS & 3.2215\\
NGC 6913 & 99 & NSVS 5731720 & 20:23:21.7 & +38:00:07 & -- & k3V: & 12.009 & 0.425 & 3.2061 & CEP & 3.2046\\
NGC 6913 & 309 & NGC 6913 36 & 20:23:41.2 & +38:27:00 & -- & f & 14.032 & 0.535 & 3.4379 & -- & --\\
NGC 7654 & 234 & PW Cas & 23:25:58.5 & +61:16:01 & -- & F8/G0 & 13.061 & 0.713 & 3.9999 & DCEP & 3.9999\\
P Cygni & 18 & TYC 3152-1813-1 & 20:20:47.1 & +38:17:07 & -- & -- & 9.987 & 0.141 & 2.5164 & CEP & 2.5137\\
P Cygni & 405 & 2MASS J20173903+3804143 & 20:17:39.0 & +38:04:14 & -- & -- & 13.591 & 0.506 & 5.0511 & -- & --\\
\end{longtable}

\begin{longtable}{lclcccccccccc}
\caption{Observed $\gamma$ Doradus variables.}\\
\label{tab:GDOR_full}\\
\hline 
\hline
\rule{0pt}{10pt}Field & ID & Designation & RA & DEC & Cluster & Spectral & $V$ & Amplitude & Period & VSX & VSX period\\
&  &  & J2000 & J2000 & membership & type & (mag) & (mag) & (days) & info  & (days)\\
\hline
\endfirsthead
\caption{continued.}\\
\hline \hline
\rule{0pt}{10pt}Field & ID & Designation & RA & DEC & Cluster & Spectral & $V$ & Amplitude & Period & VSX & VSX period\\
&  &  & J2000 & J2000 & membership & type & (mag) & (mag) & (days) & info  & (days)\\
\hline
\endhead
\hline 
\endfoot
\rule{0pt}{10pt}NGC 581 & 7 & V1132 Cas & 01:33:45.5 & +60:37:23 & -- & F1 & 8.157 & 0.097 & 0.6257 & GDOR & 0.6258\\
\end{longtable}

\begin{longtable}{lclcccccccccc}
\caption{Observed WR variables.}\\
\label{tab:WR_full}\\
\hline 
\hline
\rule{0pt}{10pt}Field & ID & Designation & RA & DEC & Cluster & Spectral & $V$ & Amplitude & Period & VSX & VSX period\\
&  &  & J2000 & J2000 & membership & type & (mag) & (mag) & (days) & info  & (days)\\
\hline
\endfirsthead
\caption{continued.}\\
\hline \hline
\rule{0pt}{10pt}Field & ID & Designation & RA & DEC & Cluster & Spectral & $V$ & Amplitude & Period & VSX & VSX period\\
&  &  & J2000 & J2000 & membership & type & (mag) & (mag) & (days) & info  & (days)\\
\hline
\endhead
\hline 
\endfoot
\rule{0pt}{10pt}Berkeley 86 & 3 & V444 Cyg & 20:19:32.4 & +38:43:54 & -- & WN5+O6II-V & 7.973 & 0.336 & 4.2130 & EA/WR & 4.2124\\
Berkeley 87 & 21 & WR 141 & 20:21:31.7 & +36:55:13 & -- & WN5-w+O5V-III & 9.840 & 0.108 & -- & WR & --\\
PZ Cas & 66 & WR 158 & 23:43:30.6 & +61:55:48 & -- & WN7 & 11.282 & 0.073 & -- & -- & --\\
\end{longtable}

\begin{longtable}{lclcccccccccc}
\caption{Observed $\alpha$ Cyg variables.}\\
\label{tab:ACYG_full}\\
\hline 
\hline
\rule{0pt}{10pt}Field & ID & Designation & RA & DEC & Cluster & Spectral & $V$ & Amplitude & Period & VSX & VSX period\\
&  &  & J2000 & J2000 & membership & type & (mag) & (mag) & (days) & info  & (days)\\
\hline
\endfirsthead
\caption{continued.}\\
\hline \hline
\rule{0pt}{10pt}Field & ID & Designation & RA & DEC & Cluster & Spectral & $V$ & Amplitude & Period & VSX & VSX period\\
&  &  & J2000 & J2000 & membership & type & (mag) & (mag) & (days) & info  & (days)\\
\hline
\endhead
\hline 
\endfoot
\rule{0pt}{10pt}Berkeley 87 & 7 & HD 229059 & 20:21:15.4 & +37:24:31 & -- & B2Iabe & 8.761 & 0.136 & 8.1319 & BE & --\\
Cygnus OB2 & 33 & Schulte 12 & 20:32:41.0 & +41:14:29 & FSR 0236 & B3.5Ia+ & 11.702 & 0.126 & -- & ? & --\\
Cygnus OB2 & 35 & Schulte 19 & 20:33:39.1 & +41:19:26 & Cygnus OB2 & B0Iab & 10.944 & 0.063 & -- & IS: & --\\
Cygnus OB2 & 43 & $[$CPR2002$]$ A23 & 20:30:39.7 & +41:08:49 & -- & B0.7Ib & 11.296 & 0.085 & -- & -- & --\\
Cygnus OB2 & 1512 & Schulte 18 & 20:33:30.8 & +41:15:23 & -- & B1Ib & 10.860 & 0.093 & -- & -- & --\\
EO Per & 7 & BD+56 733 & 02:52:40.7 & +57:27:52 & -- & B1II-III* & 9.711 & 0.048 & -- & -- & --\\
Gemini OB1 & 1 & PU Gem & 06:09:44.0 & +23:06:49 & -- & B3Ia & 5.749 & 0.102 & -- & ACYG & 6.8066\\
NGC 663 & 5 & LS I +61 232 & 01:46:23.8 & +61:15:30 & NGC 663 & B5Ib & 8.472 & 0.055 & -- & -- & --\\
\end{longtable}

\begin{longtable}{lclcccccccccc}
\caption{Observed variable Be stars.}\\
\label{tab:Be_full}\\
\hline 
\hline
\rule{0pt}{10pt}Field & ID & Designation & RA & DEC & Cluster & Spectral & $V$ & Amplitude & Period & VSX & VSX period\\
&  &  & J2000 & J2000 & membership & type & (mag) & (mag) & (days) & info  & (days)\\
\hline
\endfirsthead
\caption{continued.}\\
\hline \hline
\rule{0pt}{10pt}Field & ID & Designation & RA & DEC & Cluster & Spectral & $V$ & Amplitude & Period & VSX & VSX period\\
&  &  & J2000 & J2000 & membership & type & (mag) & (mag) & (days) & info  & (days)\\
\hline
\endhead
\hline 
\endfoot
\rule{0pt}{10pt}Berkeley 87 & 112 & VES 203 & 20:21:24.9 & +37:22:48 & -- & B0/1Ve & 12.141 & 0.194* & -- & ? & --\\
Berkeley 87 & 124 & TYC 2684-476-1 & 20:21:54.5 & +37:00:00 & -- & em & 12.226 & 0.133 & -- & -- & --\\
Berkeley 87 & 462 & AS 407 & 20:21:28.4 & +37:26:20 & -- & B2e & 14.228 & 0.356* & -- & -- & --\\
Cygnus OB2 & 109 & $[$MT91$]$ 213 & 20:32:13.1 & +41:27:24 & FSR 0236 & B0V & 11.266 & 0.534* & -- & BE & --\\
Cygnus OB2 & 121 & Schulte 30 & 20:34:43.6 & +41:29:05 & -- & B2IIIe & 12.240 & 0.220* & -- & -- & --\\
EO Per & 35 & EO Per & 02:50:38.6 & +57:20:10 & -- & B0ea & 10.938 & 0.872* & -- & ISA & --\\
IC 1805 & 44 & KM Cas & 02:29:30.5 & +61:29:44 & -- & B0III & 11.219 & 0.253* & -- & SDOR: & --\\
NGC 581 & 72 & NGC 581 2 & 01:33:24.3 & +60:39:45 & -- & B2Ve & 11.270 & 0.179 & -- & ? & --\\
\end{longtable}
\tablefoot{
Be stars with photometric outbursts are marked with an asterisk (*) behind their amplitude value.
}

\addtolength{\tabcolsep}{-2pt} 
\begin{longtable}{lclcccccccccc}
\caption{Observed variable late-type stars.}\\
\label{tab:M_full}\\
\hline 
\hline
\rule{0pt}{10pt}Field & ID & Designation & RA & DEC & Cluster & Spectral & $V$ & Amplitude & Period & VSX & VSX period\\
&  &  & J2000 & J2000 & membership & type & (mag) & (mag) & (days) & info  & (days)\\
\hline
\endfirsthead
\caption{continued.}\\
\hline \hline
\rule{0pt}{10pt}Field & ID & Designation & RA & DEC & Cluster & Spectral & $V$ & Amplitude & Period & VSX & VSX period\\
&  &  & J2000 & J2000 & membership & type & (mag) & (mag) & (days) & info  & (days)\\
\hline
\endhead
\hline 
\endfoot
\rule{0pt}{10pt}Berkeley 86 & 1 & HD 193469 & 20:18:57.5 & +39:00:14 & -- & K4.5Ib & 6.408 & 0.099 & -- & ? & --\\
Berkeley 86 & 18 & HD 228689 & 20:16:23.2 & +38:48:16 & -- & M4 & 9.792 & 0.142 & -- & -- & --\\
Berkeley 87 & 10 & BC Cyg & 20:21:38.5 & +37:31:59 & Berkeley 87 & M4I & 9.799 & 0.543 & -- & SRC & 700\\
Berkeley 87 & 12 & BI Cyg & 20:21:21.9 & +36:55:56 & Berkeley 87 & M2/4I & 9.526 & 0.401 & -- & LC & --\\
Berkeley 87 & 23 & HD 229105 & 20:21:56.2 & +37:21:29 & -- & K2II & 9.955 & 0.037 & -- & -- & --\\
EO Per & 3 & V648 Cas & 02:51:03.9 & +57:51:20 & -- & M2I: & 9.189 & 0.393 & -- & LC & --\\
EO Per & 5 & V637 Cas & 02:56:53.2 & +57:33:25 & -- & M1Ib: & 9.719 & 0.138 & -- & LC & --\\
EO Per & 253 & IRAS 02475+5704 & 02:51:13.9 & +57:16:49 & -- & C & 14.449 & 0.911 & -- & L & 193\\
\end{longtable}
\addtolength{\tabcolsep}{2pt}

\begin{longtable}{lclcccccccccc}
\caption{Observed unclassified variable stars.}\\
\label{tab:unclassified_full}\\
\hline 
\hline
\rule{0pt}{10pt}Field & ID & Designation & RA & DEC & Cluster & Spectral & $V$ & Amplitude & Period & VSX & VSX period\\
&  &  & J2000 & J2000 & membership & type & (mag) & (mag) & (days) & info  & (days)\\
\hline
\endfirsthead
\caption{continued.}\\
\hline \hline
\rule{0pt}{10pt}Field & ID & Designation & RA & DEC & Cluster & Spectral & $V$ & Amplitude & Period & VSX & VSX period\\
&  &  & J2000 & J2000 & membership & type & (mag) & (mag) & (days) & info  & (days)\\
\hline
\endhead
\hline 
\endfoot
\rule{0pt}{10pt}Berkeley 86 & 7 & HD 228898 & 20:19:17.2 & +38:25:50 & -- & K5 & 9.169 & 0.085 & -- & -- & --\\
Berkeley 86 & 36 & HD 228905 & 20:19:21.3 & +38:24:41 & Berkeley 86 & OB- & 10.229 & 0.125 & -- & -- & --\\
Berkeley 87 & 4 & HD 194709 & 20:25:42.1 & +37:08:43 & -- & K2 & 8.387 & 0.036 & -- & -- & --\\
Berkeley 87 & 24 & BD+37 3927 & 20:25:06.9 & +37:27:56 & -- & O8.5I & 10.177 & 0.051 & -- & -- & --\\
Berkeley 87 & 26 & LS II +37 83 & 20:22:58.3 & +37:14:05 & -- & OB & 10.169 & 0.036 & -- & -- & --\\
Berkeley 87 & 35 & LS II +36 66 & 20:21:37.9 & +37:08:15 & -- & OB & 10.621 & 0.042 & -- & -- & --\\
Berkeley 87 & 41 & TYC 2684-1267-1 & 20:20:56.3 & +37:07:32 & -- & K5V* & 11.189 & 0.071 & -- & -- & --\\
Berkeley 87 & 49 & HD 229142 & 20:22:32.2 & +36:55:55 & -- & A & 10.603 & 0.040 & -- & -- & --\\
\end{longtable}

\begin{longtable}{lclcccccccc}
\caption{Variable stars not in VSX.}\\
\label{tab:new_full}\\
\hline 
\hline
\rule{0pt}{10pt}Field & ID & Designation & RA & DEC & Cluster & Spectral & $V$ & Amplitude & Period & Variability\\
&  &  & J2000 & J2000 & membership & type & (mag) & (mag) & (days) & type  \\
\hline
\endfirsthead
\caption{continued.}\\
\hline \hline
\rule{0pt}{10pt}Field & ID & Designation & RA & DEC & Cluster & Spectral & $V$ & Amplitude & Period & Variability\\
&  &  & J2000 & J2000 & membership & type & (mag) & (mag) & (days) & type\\
\hline
\endhead
\hline 
\endfoot
\rule{0pt}{10pt}Berkeley 86 & 7 & HD 228898 & 20:19:17.2 & +38:25:50 & -- & K5 & 9.169 & 0.085 & -- & --\\
Berkeley 86 & 18 & HD 228689 & 20:16:23.2 & +38:48:16 & -- & M4 & 9.792 & 0.142 & -- & SR\\
Berkeley 86 & 22 & HD 228989 & 20:20:21.4 & +38:42:00 & -- & O8.5V:+O9.7V: & 9.686 & 0.091 & 1.7734 & E\\
Berkeley 86 & 28 & LS II +38 39 & 20:18:59.4 & +38:57:14 & -- & B1III & 10.057 & 0.179 & 0.2286 & BCEP\\
Berkeley 86 & 36 & HD 228905 & 20:19:21.3 & +38:24:41 & Berkeley 86 & OB- & 10.229 & 0.125 & -- & --\\
Berkeley 86 & 52 & LS II +38 43 & 20:19:36.8 & +38:35:21 & Berkeley 86 & B1V & 10.676 & 0.148 & 1.9102 & E\\
Berkeley 86 & 323 & 2MASS J20194609+3851325 & 20:19:46.1 & +38:51:32 & -- & -- & 12.855 & 0.372 & 0.1381 & BCEP\\
Berkeley 86 & 381 & 2MASS J20190590+3853009 & 20:19:05.9 & +38:53:01 & -- & -- & 12.808 & 0.569 & 4.3262 & E\\
\hline 
\end{longtable}
\tablefoot{
Variability types: E - Eclipsing binary; BE - Be star; BCEP - $\beta$ Cephei type star; SPB - Slowly pulsating B star; ACYG - $\alpha$ Cygni type star; CEP - Cepheid; WR - Wolf-Rayet star; SR - late-type star; "--" - unclassified.
}
\end{landscape}

\section{Individual fields}\label{sec:app_field}

\begin{figure*}
        \includegraphics[width=0.89\textwidth, trim=0 0 0 0,clip]{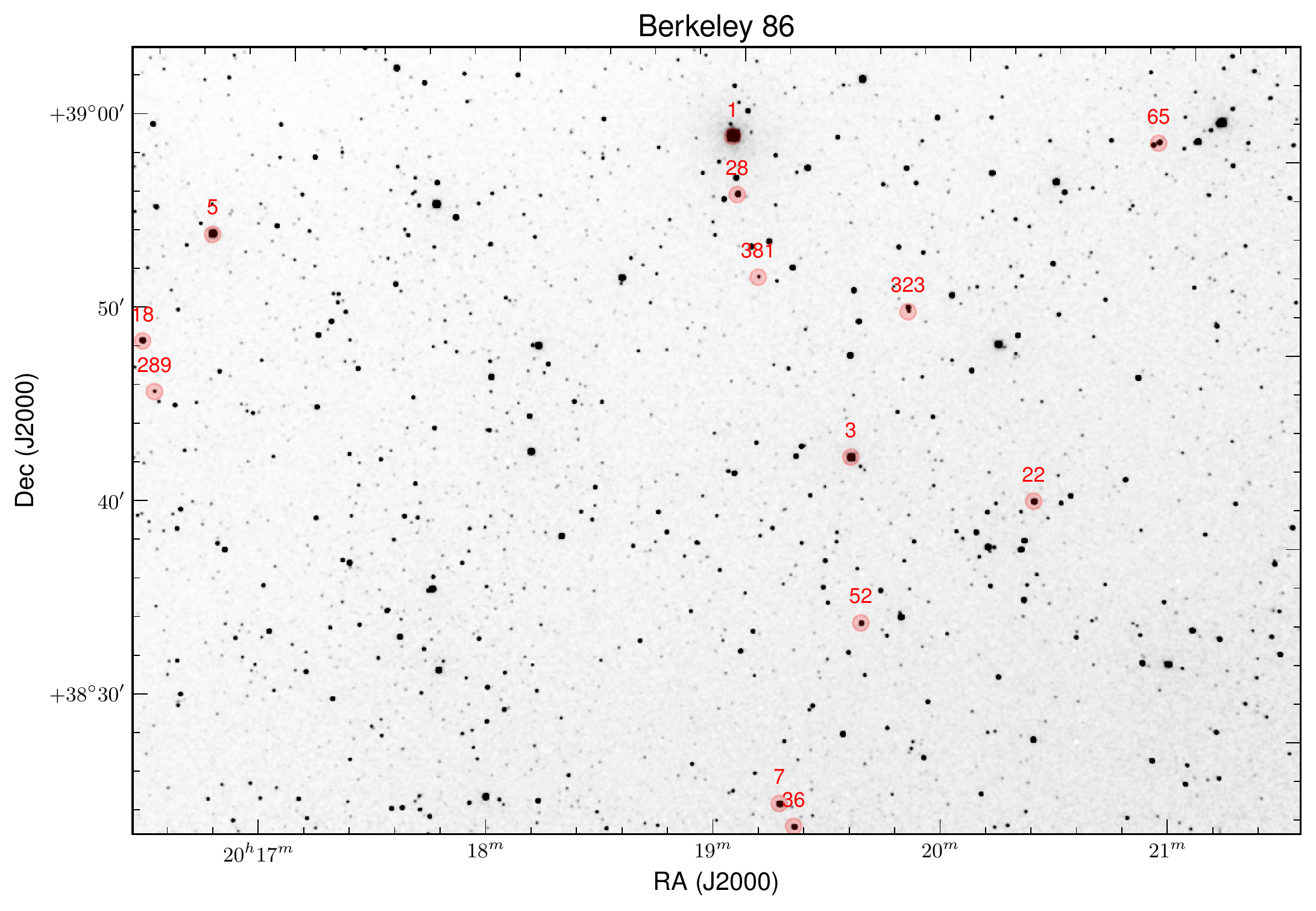}
        \centering
        \caption{Variable stars in Berkeley 86 field. Variable stars are marked in red and the ID numbers can be cross-referenced from the second column of the Tables presented in Appendix \ref{sec:app_variables}.}
        \label{field-Berk86}
\end{figure*}

\begin{figure*}
        \includegraphics[width=0.9\textwidth, trim=0 0 0 0,clip]{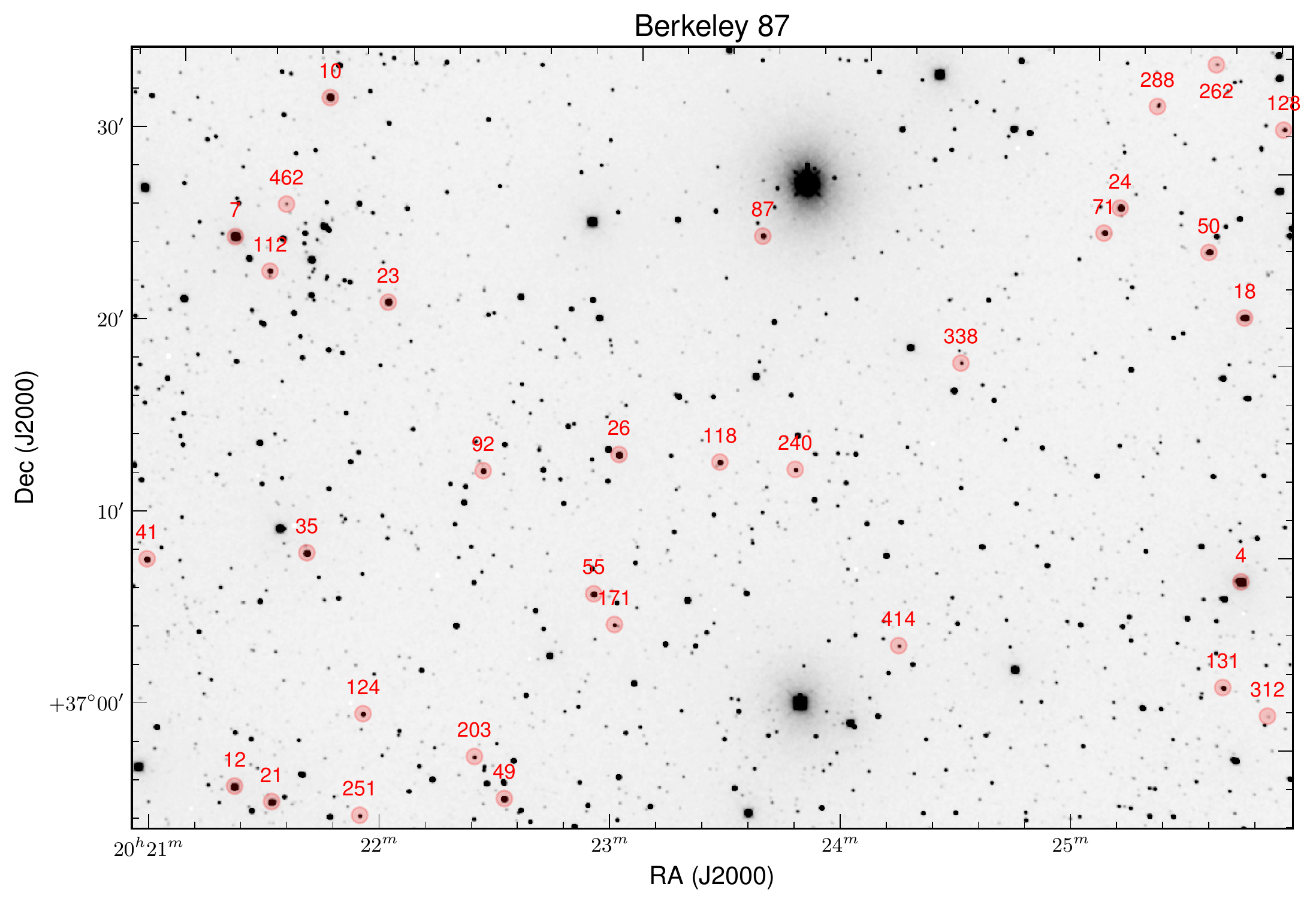}
        \centering
        \caption{Variable stars in Berkeley 87 field}
        \label{field-Berk87}
\end{figure*}

\begin{figure*}
        \includegraphics[width=0.9\textwidth, trim=0 0 0 0,clip]{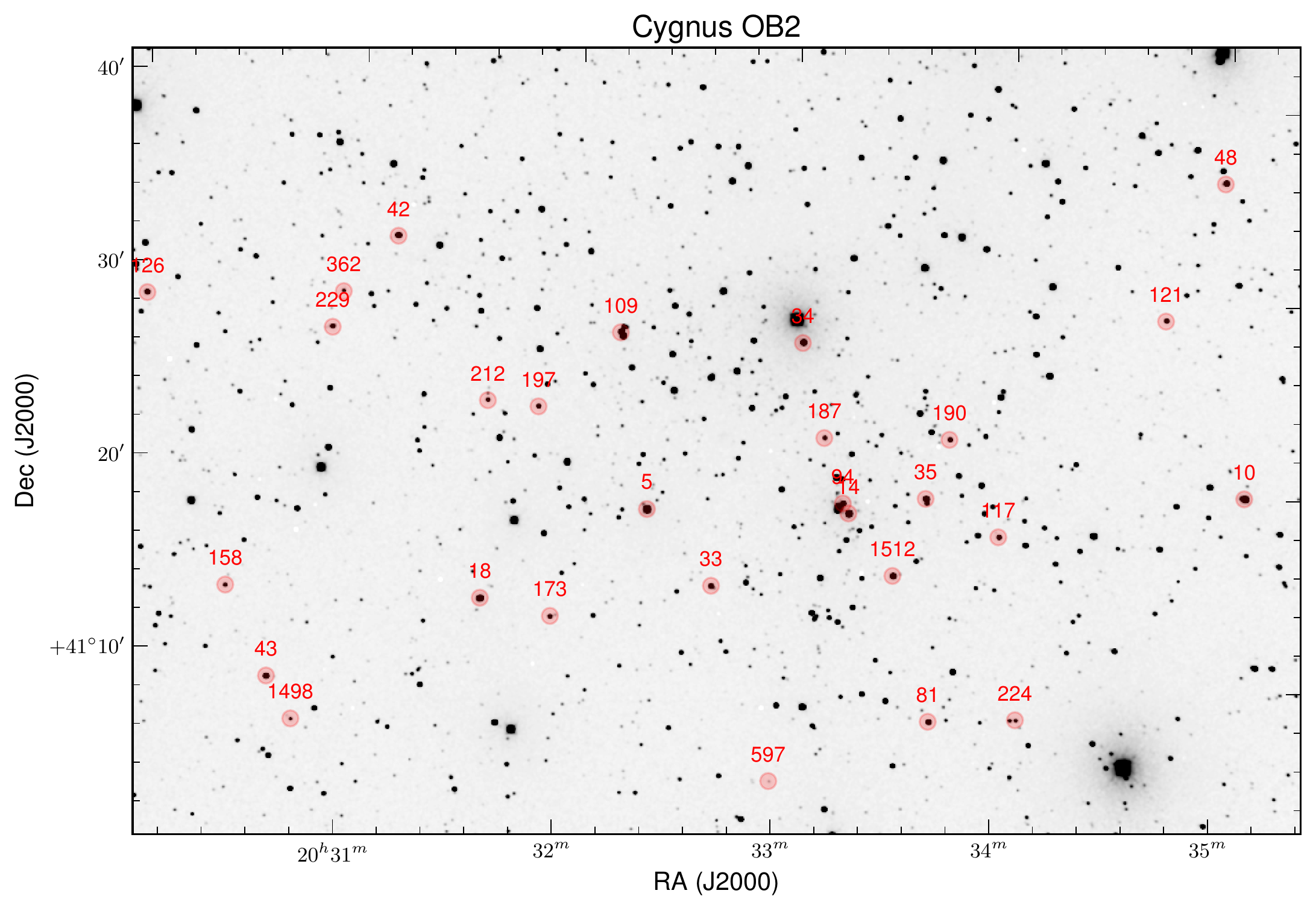}
        \centering
        \caption{Variable stars in Cygnus OB2 field}
        \label{field-CygOB2}
\end{figure*}

\begin{figure*}
        \includegraphics[width=0.9\textwidth, trim=0 0 0 0,clip]{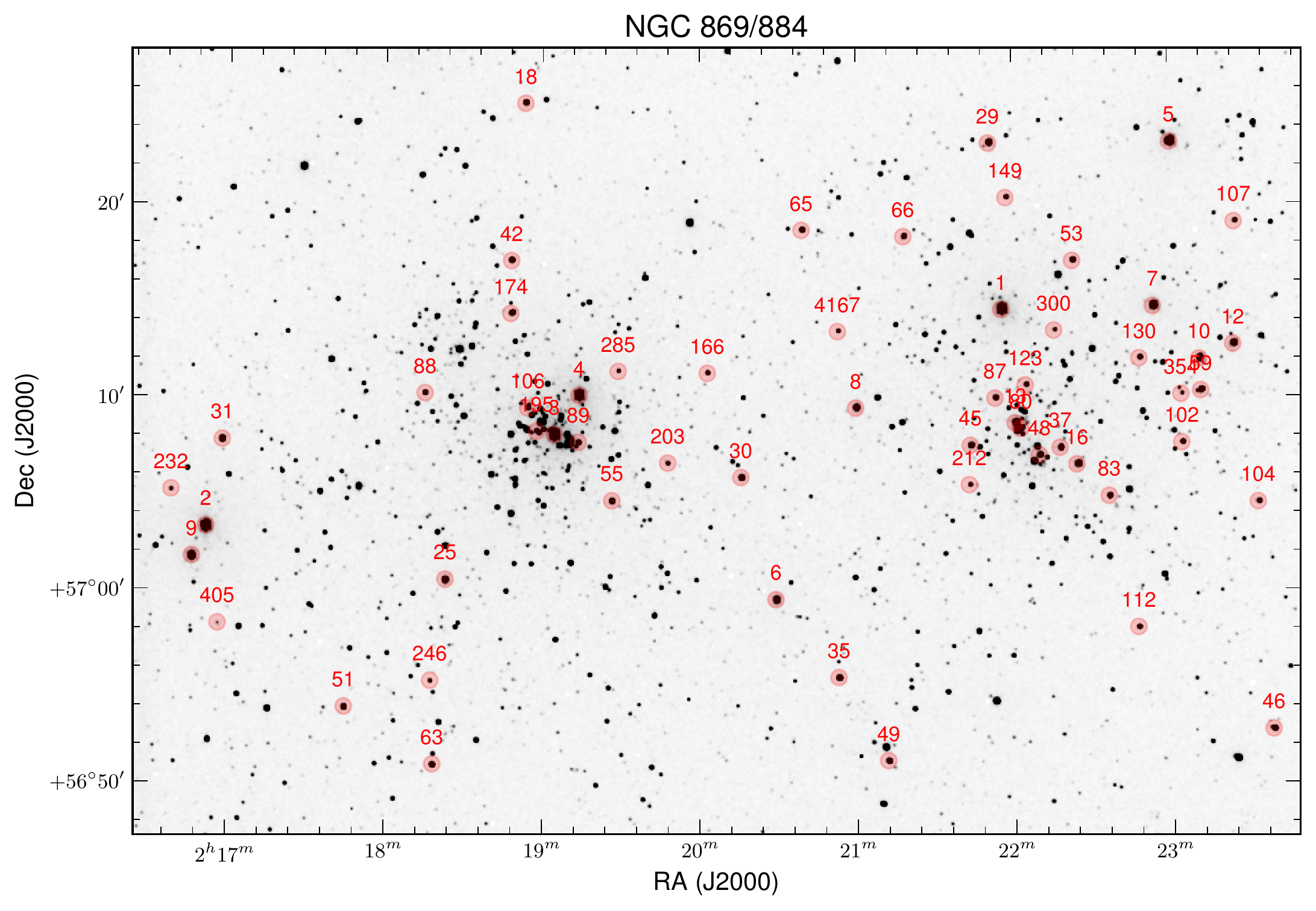}
        \centering
        \caption{Variable stars in NGC 869/884 field}
        \label{field-Double}
\end{figure*}

\begin{figure*}
        \includegraphics[width=0.9\textwidth, trim=0 0 0 0,clip]{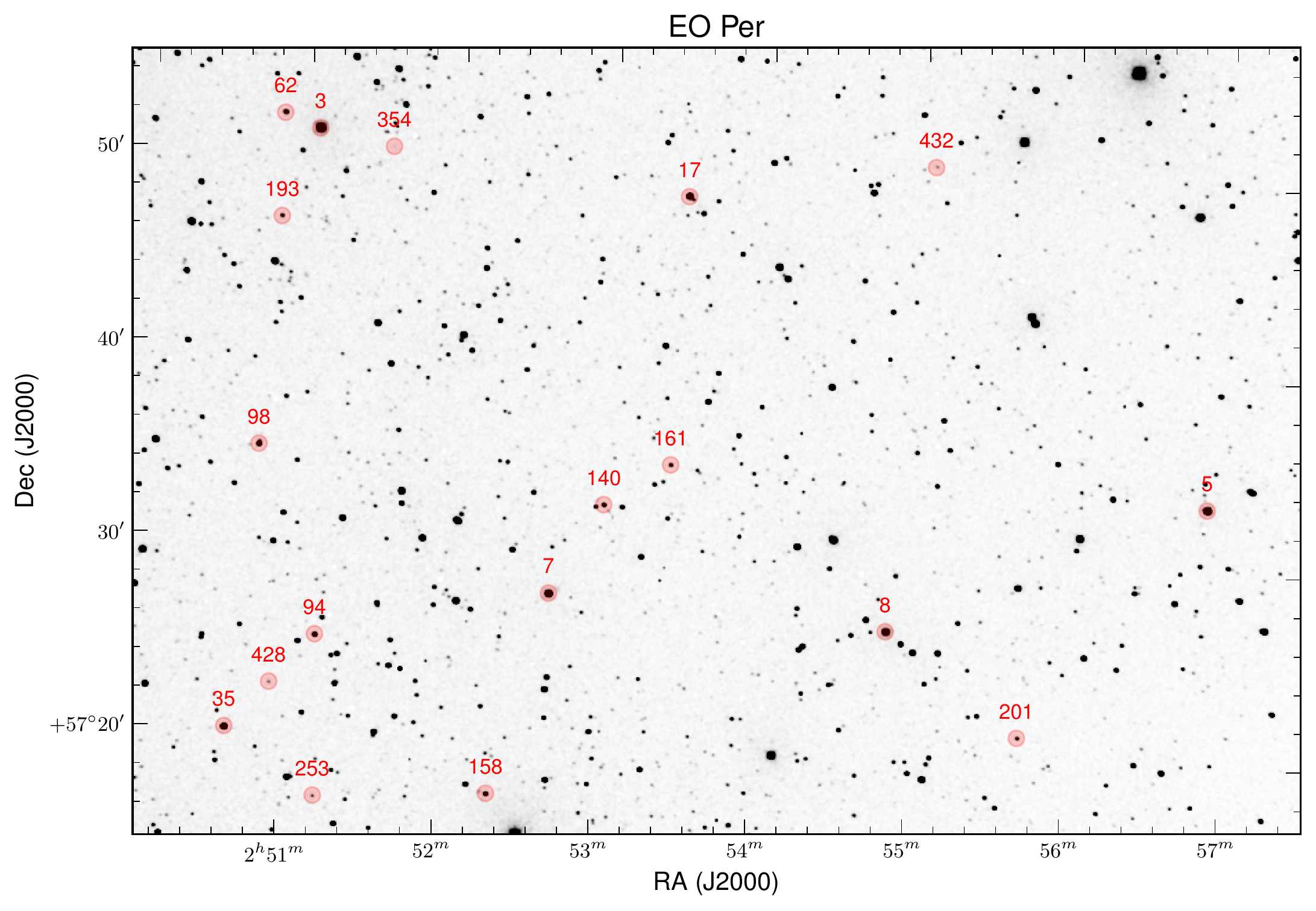}
        \centering
        \caption{Variable stars in EO Per field}
        \label{field-EOPer}
\end{figure*}

\begin{figure*}
        \includegraphics[width=0.9\textwidth, trim=0 0 0 0,clip]{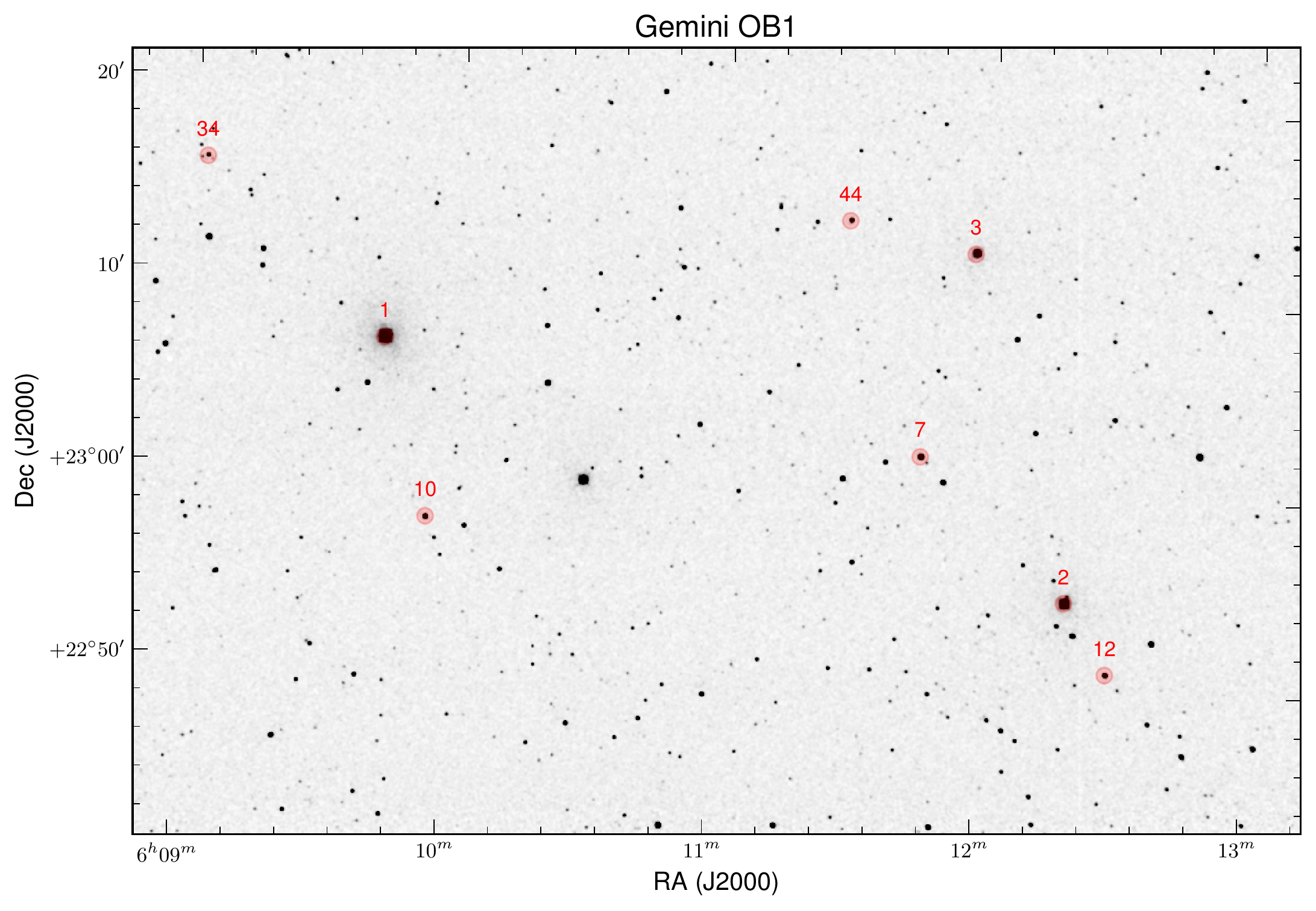}
        \centering
        \caption{Variable stars in Gemini OB1 field}
        \label{field-GemOB1}
\end{figure*}

\begin{figure*}
        \includegraphics[width=0.9\textwidth, trim=0 0 0 0,clip]{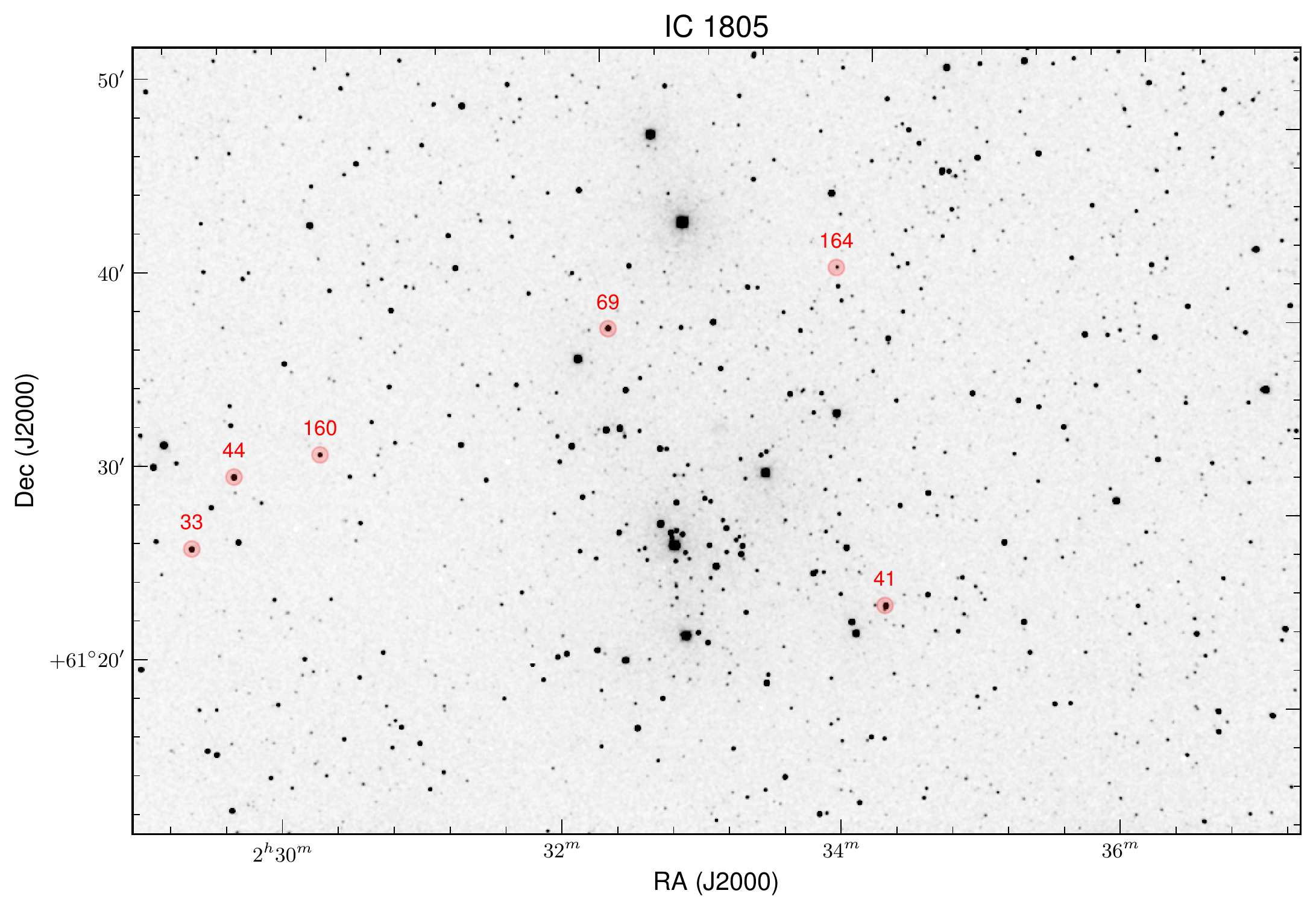}
        \centering
        \caption{Variable stars in IC 1805 field}
        \label{field-IC1805}
\end{figure*}

\begin{figure*}
        \includegraphics[width=0.9\textwidth, trim=0 0 0 0,clip]{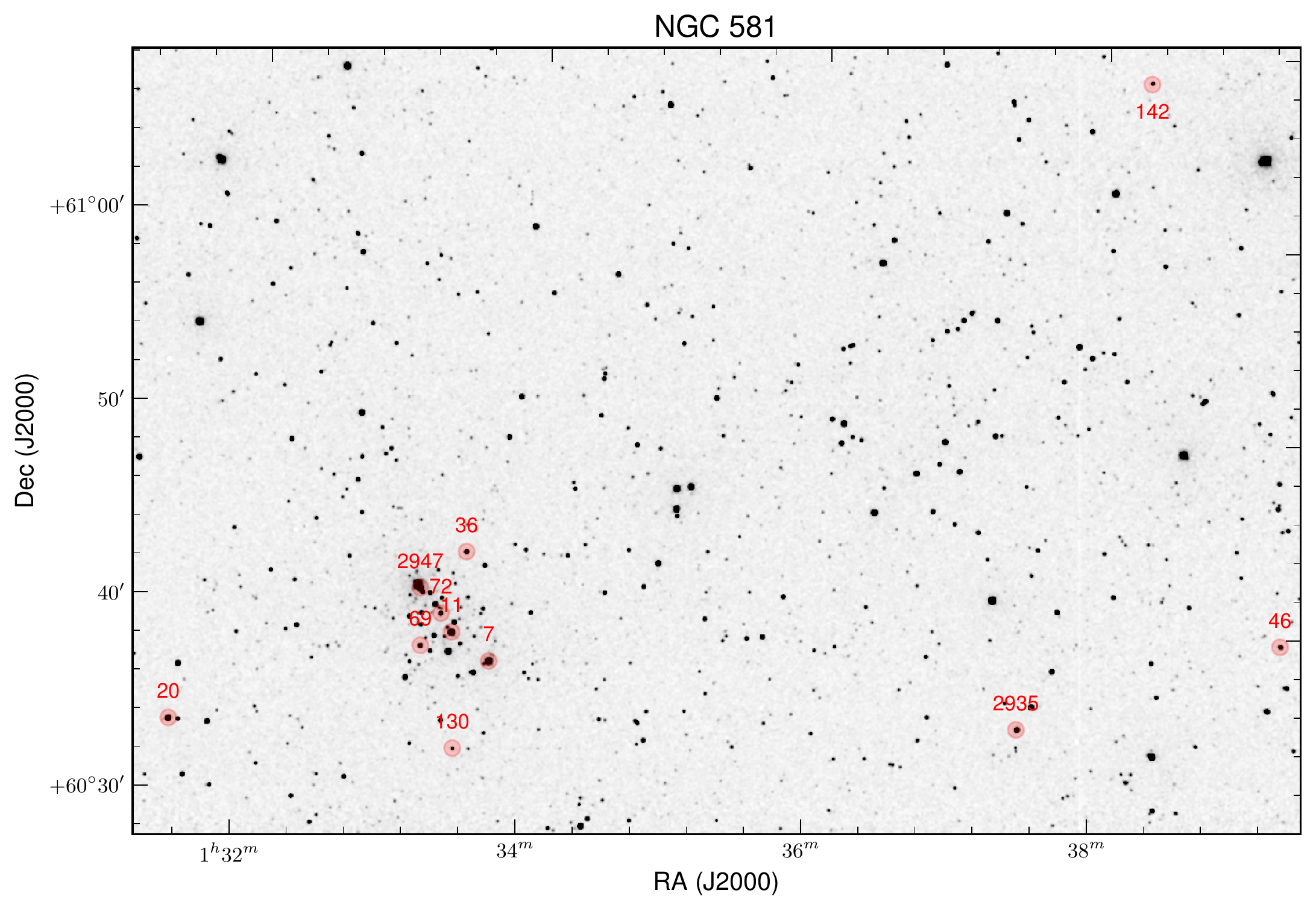}
        \centering
        \caption{Variable stars in NGC581 field}
        \label{field-NGC581}
\end{figure*}

\begin{figure*}
        \includegraphics[width=0.9\textwidth, trim=0 0 0 0,clip]{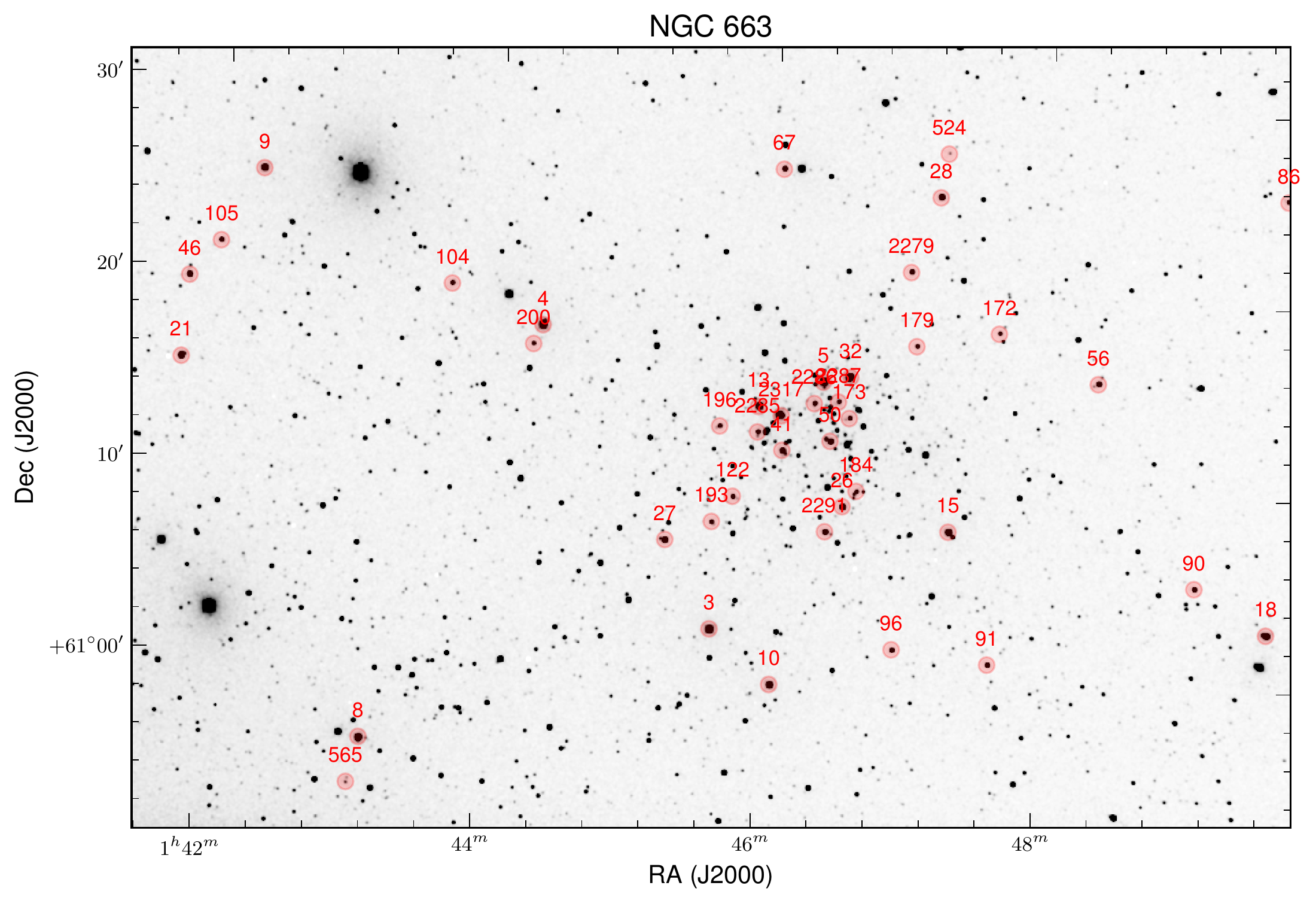}
        \centering
        \caption{Variable stars in NGC663 field}
        \label{field-NGC663}
\end{figure*}

\begin{figure*}[t]
        \includegraphics[width=0.9\textwidth, trim=0 0 0 0,clip]{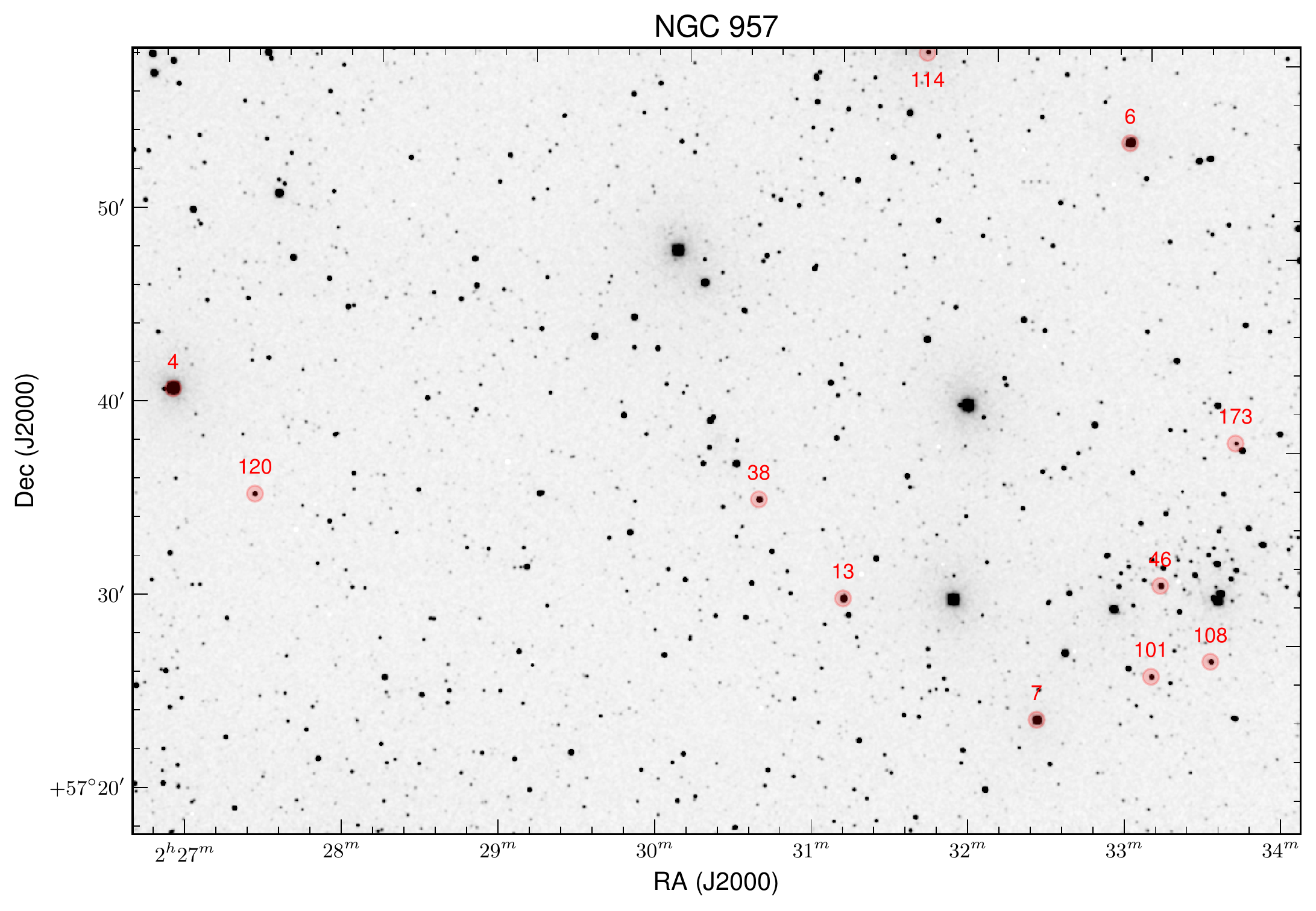}
        \centering
        \caption{Variable stars in NGC957 field}
        \label{field-NGC957}
\end{figure*}

\begin{figure*}
        \includegraphics[width=0.9\textwidth, trim=0 0 0 0,clip]{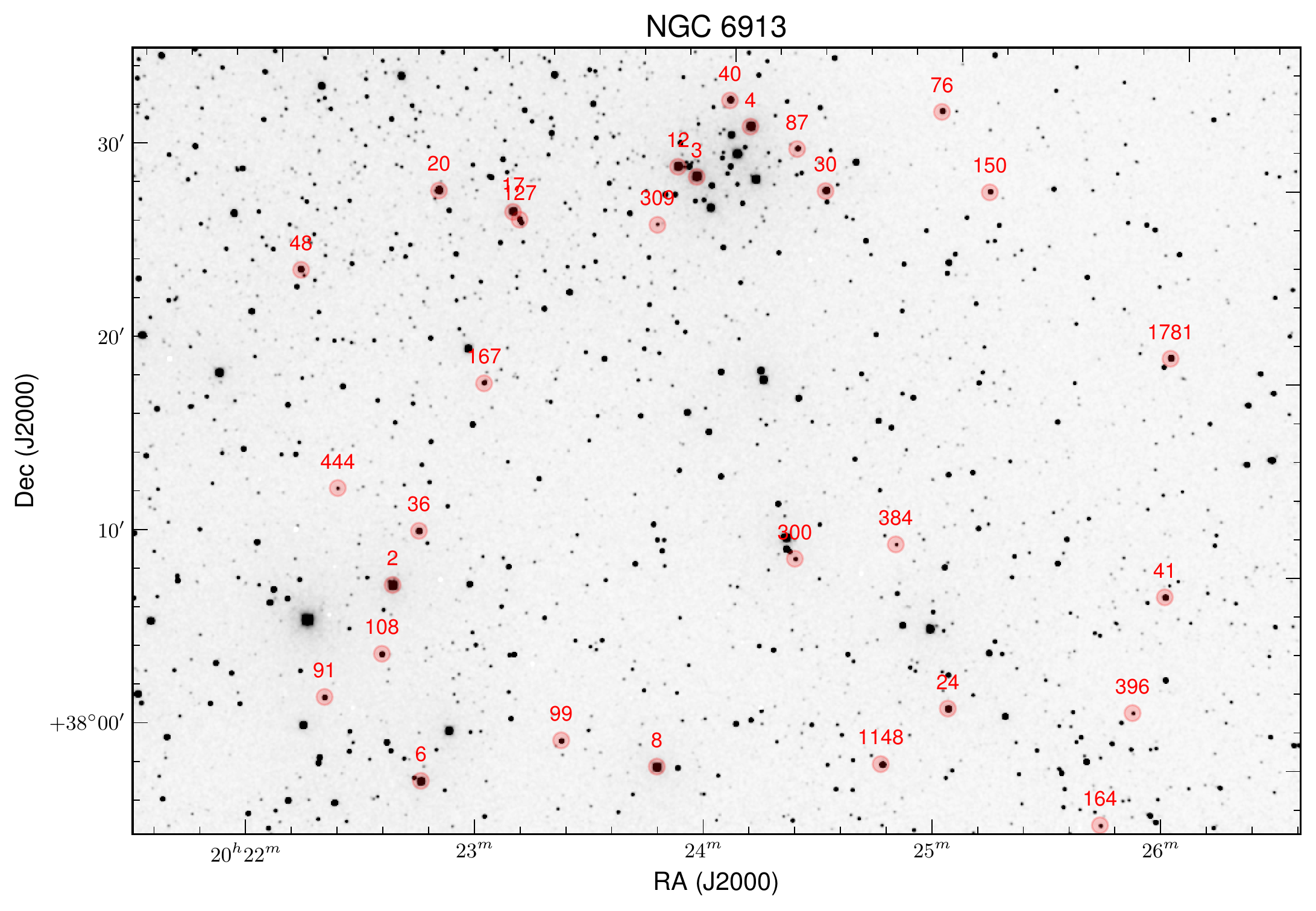}
        \centering
        \caption{Variable stars in NGC6913 field}
        \label{field-NGC6913}
\end{figure*}

\begin{figure*}
        \includegraphics[width=0.9\textwidth, trim=0 0 0 0,clip]{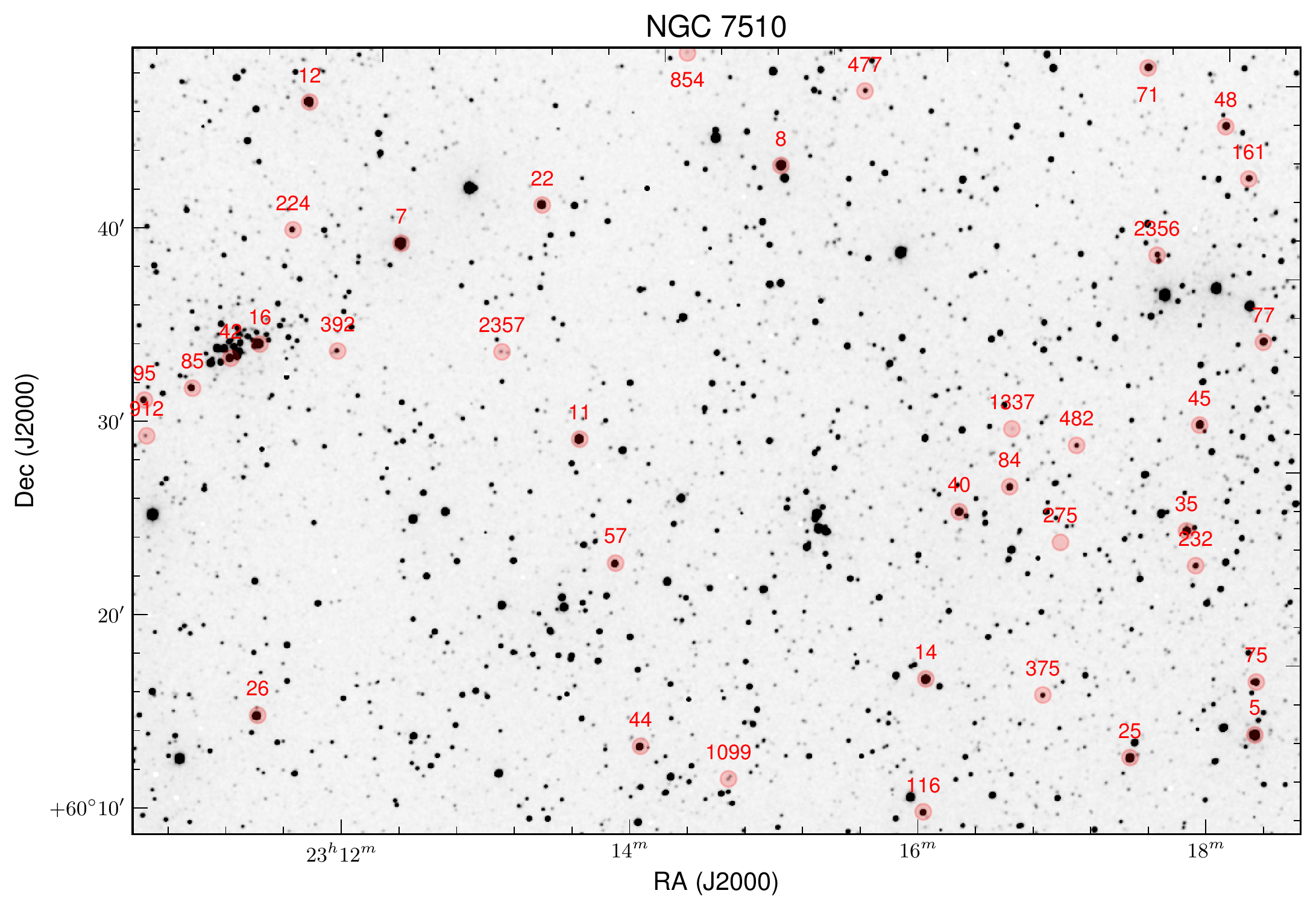}
        \centering
        \caption{Variable stars in NGC7510 field}
        \label{field-NGC7510}
\end{figure*}

\begin{figure*}
        \includegraphics[width=0.9\textwidth, trim=0 0 0 0,clip]{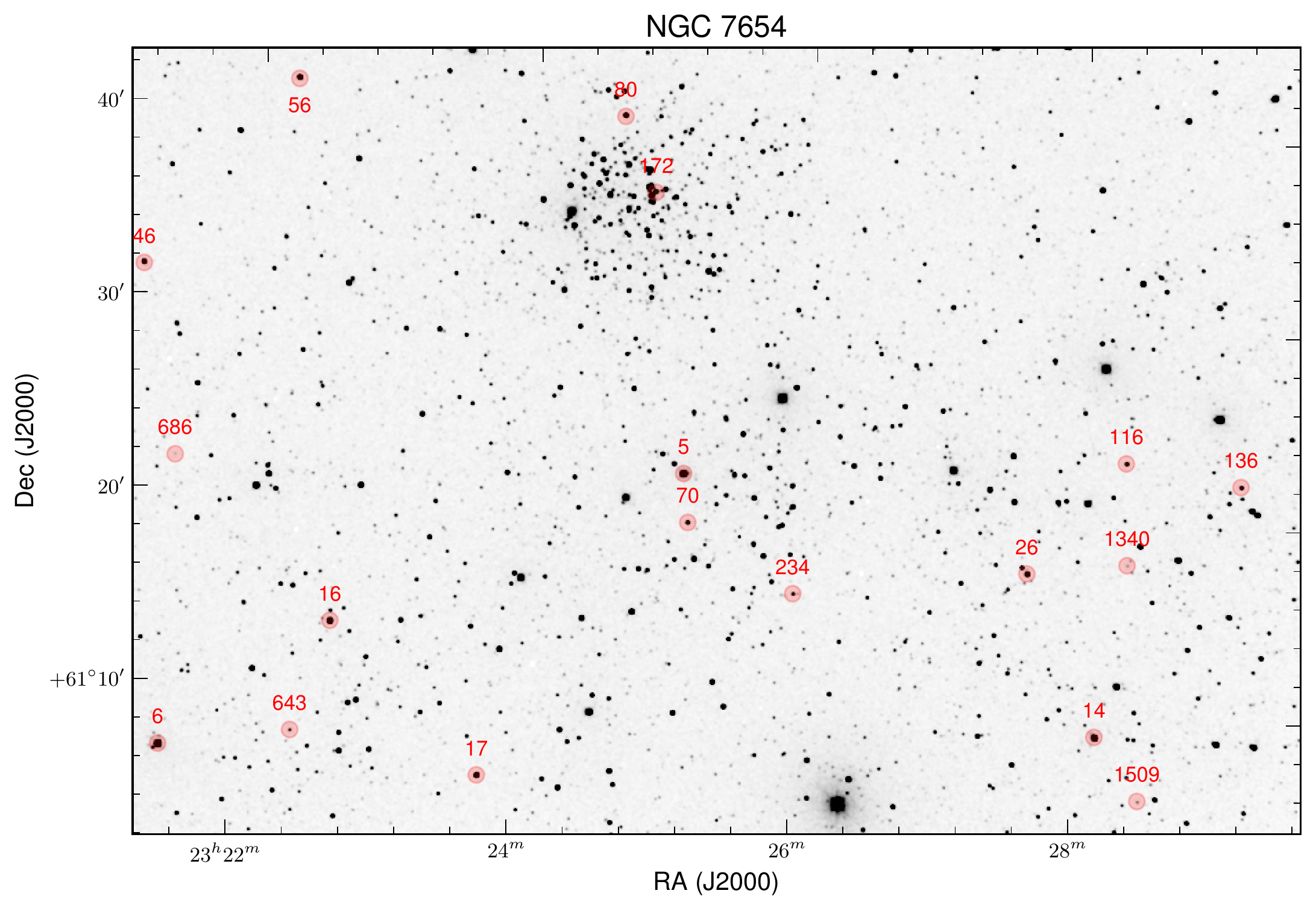}
        \centering
        \caption{Variable stars in NGC7654 field}
        \label{field-NGC7654}
\end{figure*}

\begin{figure*}
        \includegraphics[width=0.9\textwidth, trim=0 0 0 0,clip]{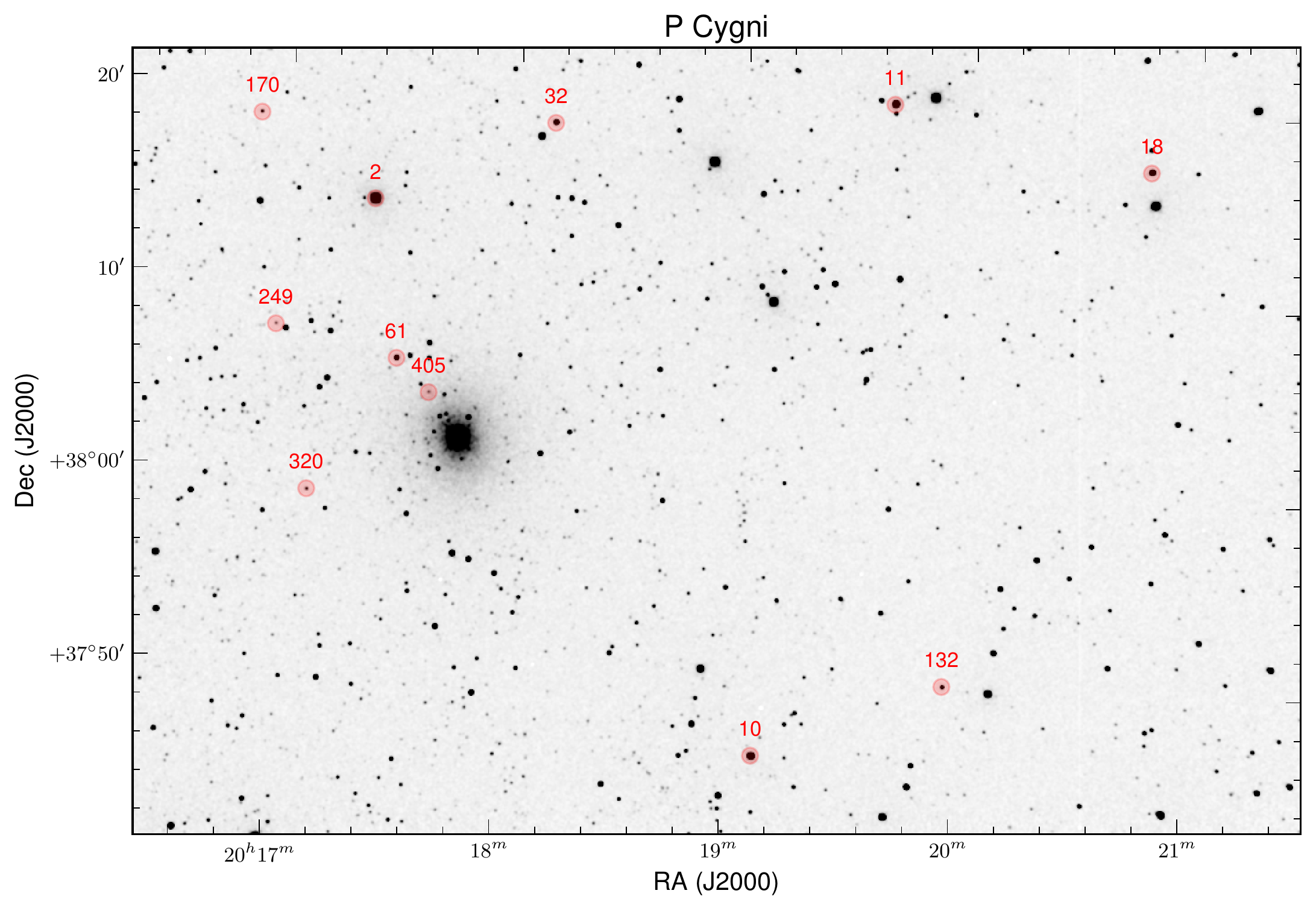}
        \centering
        \caption{Variable stars in P Cygni field}
        \label{field-PCyg}
\end{figure*}

\begin{figure*}
        \includegraphics[width=0.9\textwidth, trim=0 0 0 0,clip]{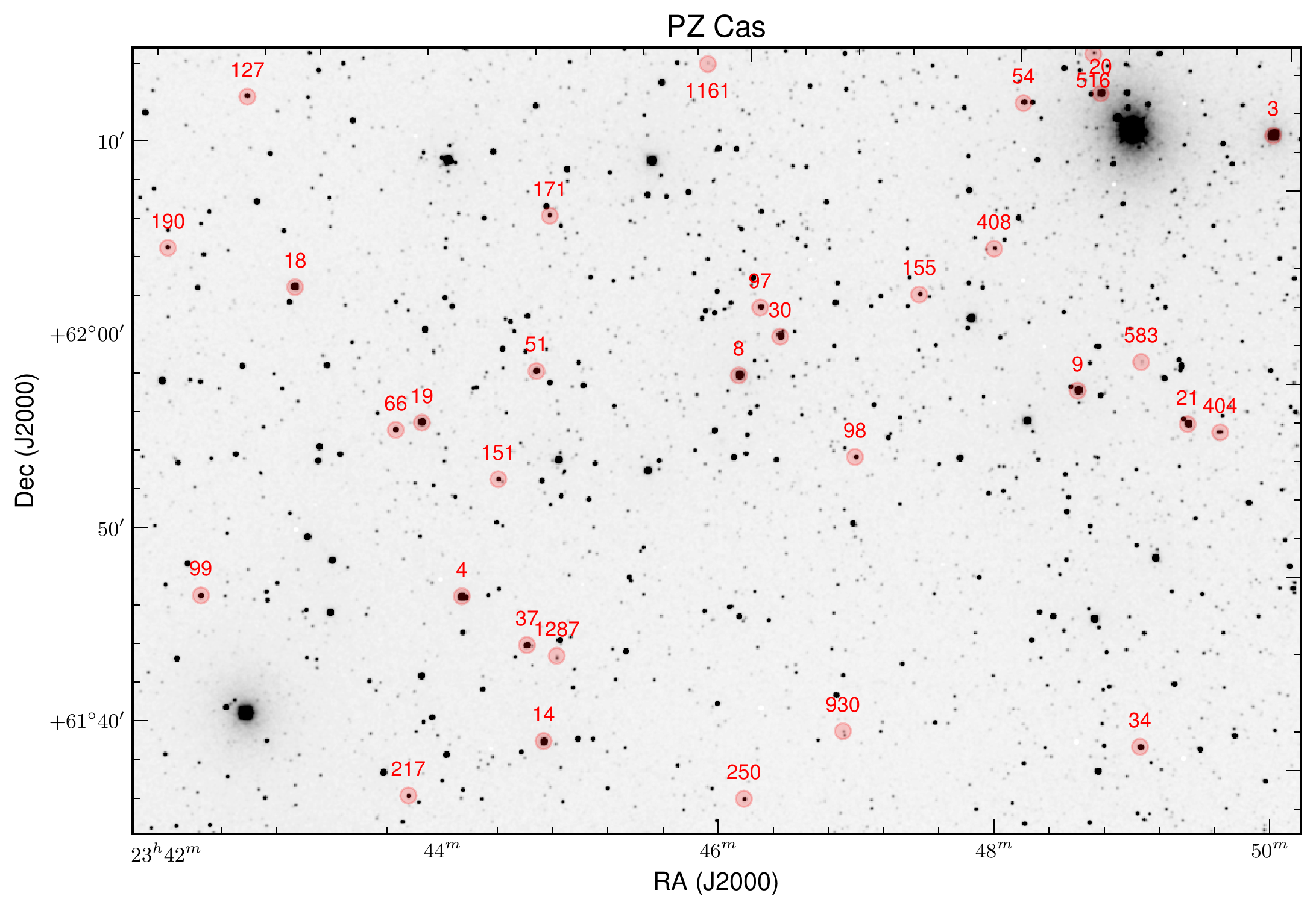}
        \centering
        \caption{Variable stars in PZ Cas field}
        \label{field-PZCas}
\end{figure*}

We present finding charts for our variable star candidates across the 15 fields in Fig. \ref{field-Berk86} -- \ref{field-PZCas}.
These finding charts were obtained by median smoothing our observed $V$ passband frames taken with the 0.25-m Takahashi Epsilon telescope, located in Mayhill, New Mexico, USA.
All of the frames were with a fixed field of view of $60.5~\times~40.8$ arcmin with pixel scale of 1.64 arcsec.
Variable stars are indicated with a red circle and our internal ID number above them.
We have only marked the stars listed in our variable star catalogues in Appendix \ref{sec:app_variables} and the IDs can be cross-referenced from the second column.

\end{appendix}
\end{document}